\documentclass[11pt,onecolumn]{IEEEtran}

\newif\ifthreevalue

\usepackage[utf8]{inputenc} 
\usepackage[T1]{fontenc}
\usepackage{url}              

\usepackage[cmex10]{amsmath}  
\usepackage{mleftright}       
\mleftright                   

\usepackage{graphicx}         
\usepackage{booktabs}         

\IEEEoverridecommandlockouts
\usepackage{amsmath,amssymb,amsfonts,amsthm}
\usepackage{enumitem}
\usepackage{mathtools}
\usepackage{algorithmic}
\usepackage{textcomp}
\usepackage[table,xcdraw]{xcolor}
\usepackage{tikz}

\usepackage{subcaption}
\usepackage{multicol,multirow}
\usepackage[style=ieee,citestyle=numeric-comp,sorting=nty]{biblatex}
\newtheorem{theorem}{Theorem}
\newtheorem{remark}{Remark}

\newtheorem{proposition}{Proposition}
\newtheorem{example}{Example}
\newtheorem{lemma}{Lemma}
\newtheorem{definition}{Definition}

\newtheorem{corollary}{Corollary}
\newtheorem{observation}{Observation}

\DeclareMathOperator*{\smallplus}{\scalerel*{+}{\textstyle\sum}}
\usepackage{scalerel}

\newcommand{\cir}[1]{\tikz[baseline]{%
		\node[anchor=base, draw, circle, inner sep=0, minimum width=1.2em]{#1};}}

\newcommand{\bF}{\mathbb{F}}

\newcommand{\bN}{\mathbb{N}}

\newcommand{\bR}{\mathbb{R}}

\newcommand{\cA}{\mathcal{A}}
\newcommand{\cB}{\mathcal{B}}
\newcommand{\cC}{\mathcal{C}}
\newcommand{\cD}{\mathcal{D}}

\newcommand{\cF}{\mathcal{F}}

\newcommand{\boldb}{\mathbf{b}}
\newcommand{\boldc}{\mathbf{c}}

\newcommand{\boldg}{\mathbf{g}}

\newcommand{\bolds}{\mathbf{s}}

\newcommand{\boldv}{\mathbf{v}}
\newcommand{\boldw}{\mathbf{w}}
\newcommand{\boldx}{\mathbf{x}}
\newcommand{\boldy}{\mathbf{y}}

\DeclareSymbolFont{bbold}{U}{bbold}{m}{n}
\DeclareSymbolFontAlphabet{\mathbbold}{bbold}
\newcommand{\1}{\mathbbold{1}}
\begin{document}
	
	\title{Access-Redundancy Tradeoffs in\\Quantized Linear Computations}
	
	\author{Vinayak Ramkumar,  Netanel Raviv, and Itzhak Tamo\\
		\thanks{
  This work was presented in part at the IEEE International Symposium on Information Theory (ISIT) 2023, and at the Allerton Conference on Communication, Control, and Computing 2023. 

  Vinayak Ramkumar is jointly with the  Department of Electrical Engineering--Systems, Tel Aviv University, Tel Aviv, Israel and the Department of Computer Science and Engineering, Washington University in St. Louis, St. Louis, MO, USA. Netanel Raviv is with the Department of Computer Science and Engineering, Washington University in St. Louis, St. Louis, MO, USA. Itzhak Tamo is with the Department of Electrical Engineering--Systems, Tel Aviv University, Tel Aviv, Israel. e-mail:	vinram93@gmail.com, netanel.raviv@wustl.edu, tamo@tauex.tau.ac.il. 
  
  The work of Vinayak Ramkumar and Itzhak Tamo was supported by the European Research Council (ERC grant number 852953).}
	}
	\maketitle

	\begin{abstract}
	Linear real-valued computations over distributed datasets are common in many applications, most notably as part of machine learning inference. In particular, linear computations that are quantized, i.e., where the coefficients are restricted to a predetermined set of values (such as~$\pm 1$), have gained increasing interest lately due to their role in efficient, robust, or private machine learning models. Given a dataset to store in a distributed system, we wish to encode it so that all such computations could be conducted by accessing a small number of servers, called the \textit{access parameter} of the system. Doing so relieves the remaining servers to execute other tasks. Minimizing the access parameter gives rise to an \textit{access-redundancy} tradeoff, where a smaller access parameter requires more redundancy in the system, and vice versa.	In this paper, we study this tradeoff and provide several explicit low-access schemes for $\{\pm1\}$ quantized linear computations based on covering codes in a novel way. While the connection to covering codes has been observed in the past, our results strictly outperform the state-of-the-art for two-valued linear computations. We further show that the same storage scheme can be used to retrieve any linear combination with two distinct coefficients---regardless of what those coefficients are---with the same access parameter. This universality result is then extended to all possible quantizations with any number of values; while the storage  remains identical, the access parameter increases according to a new additive-combinatorics property we call \textit{coefficient complexity}. We then turn to study the coefficient complexity---we characterize the complexity of small sets of coefficients, provide bounds, and identify coefficient sets having the highest and lowest complexity. Interestingly, arithmetic progressions have the lowest possible complexity, and some geometric progressions have the highest possible complexity, the former being particularly attractive for its common use in uniform quantization. 
	\end{abstract}
	
	\pagestyle{plain}
	
	\section{Introduction}
A typical distributed computation setup contains a central resource-limited entity, called user, that holds data and wishes to delegate storage and computation thereof to servers in a distributed system. At a future point in time, the user wishes to conduct a certain computation over the distributed data, contacts a certain number of servers with queries, and combines their responses in order to finalize the computation. While the computation family is known at the time of storage (say, polynomials), the particular computation is not. The above setup gives rise to various challenges, such as straggling servers, adversaries, and privacy. The practice of adding redundancy to the data in order to address those challenges is commonly referred to as \textit{coded computation}~\cite{TandonLDK17, yu2017polynomial,yu2019lagrange}. Virtually all coded computation literature relies on a basic assumption: in computation time, the user contacts \textit{all} servers. As a result, any computation query can potentially occupy all the system’s resources, some of which might end up being discarded due to the straggler effect. Preemptively reducing the number of contacted servers frees up the remaining ones to serve other queries or conduct different tasks in the interim.
	
 This paper addresses the design of coding methodologies that require the central server (``user'') to contact only a fraction of the storage servers (``nodes'') for any given computation. This strategy optimizes the fraction of the system that becomes occupied by the current query and frees the remaining nodes to handle other tasks or serve other users.
At storage time, the user may use coding to add redundancy to the stored data so that any computation from the predetermined family can be computed by accessing fewer servers than what would have been required to download the data in its entirety. A clear tradeoff emerges---for any (finite) computation family one can reduce access dramatically at the price of high redundancy, by performing all computations at storage time (max. redundancy min. access). On the other hand, the user can employ general-purpose storage, obtain the data whenever needed, and conduct the computation locally (min. redundancy max. access). Formally, given a family~$\cF$ of computations of interest, and an~\textit{access} parameter~$\ell$, data is stored so that at any point in the future, any function~$f\in\cF$ can be computed by accessing at most~$\ell$ nodes. 
By and large, the purpose of this area of study is to characterize all the intermediate points in this tradeoff. That is, one wishes to characterize the feasible region (or the Pareto optimal front) of the set of all feasible access-redundancy pairs.  
For linear computations over finite fields, this problem is equivalent to the existence of linear covering codes (a folklore result, see~\cite{cohen1997covering}). 

In this paper, we focus on linear computations over~$\bR$, whose coefficients are quantized to a finite set of values (e.g.,~$\{\pm1\}$). Such computations have gained increasing attention of late, mostly for applications relating to machine learning inference, in which they have proven beneficial in terms of robustness~\cite{qin2021towards,jia2020efficient,raviv2021enhancing} and privacy~\cite{raviv2022information}.
The problem of quantized computations over~$\bR$ has been studied in the past. Motivated by applications in On-Line Analytical Processing (OLAP), the authors of \cite{ho1998partial} showed that for real-valued linear computations with~$\{0,1\}$ coefficients, binary covering codes could also be used (yet in a very different way than the finite field case), to achieve rather attractive points on the access-redundancy tradeoff. 

\subsection*{Our Contributions}
\begin{itemize} [leftmargin=*]
\item For two-valued linear computations (i.e., where the vector of coefficients contains at most two different values), our techniques rely on binary covering codes in a novel way and require covering codes with certain closure properties. By constructing such covering codes using existing tools, our results outperform the results in~\cite{ho1998partial}. Furthermore, we show that for the computation family of all real-valued linear combinations that contain only two distinct coefficients, the access-redundancy tradeoff is \textit{identical} to that of linear computations with any two particular coefficients (say~$\{\pm1\}$). This observation leads to partial \textit{universality} of two-valued computation protocols. One can store the data using, say, a scheme tailored for~$\{\pm1\}$ coefficients, and at a later point in time retrieve any linear combination with coefficients in~$\{a,b\}$, for any~$a,b\in\bR$, by accessing the same number of servers as in the~$\{\pm1\}$ scheme (to be precise, at most one additional server may be needed).

\item We then take this universality result several steps further. We show that one can store the data according to a scheme tailored for any two values (say~$\{\pm1\}$), and at a later point in time, retrieve \textit{any} linear combination with \textit{any} finite number of distinct coefficients. While the storage remains unchanged, the access parameter increases as a function of a property we call the \textit{coefficient complexity}. This result is particularly attractive since it enables the user to store the data without any knowledge of the required future coefficients; the user is free to choose both the number and the structure of those coefficients at computation time (not at storage time), yet ideally, aim for coefficients of low complexity. This complexity definition uses standard additive-combinatorics notions, but to the best of our knowledge was not studied before. 

\item After showing the implication of this complexity notion to low-access quantized computation, we proceed to study its properties. We characterize the complexity of all coefficient sets of size~$3$ and~$4$, provide lower and upper bounds, and present several common cases in which those are attained. Notably, sets that form an arithmetic progression attain the lowest possible complexity (logarithmic in their size), and hence have low access parameters; these sets appear in uniform quantization~\cite[Chapter~3]{gallager2008principles}, a popular method for simple quantization of continuous sources. On the other hand, geometric progressions have the highest possible complexity (equal to their size minus one), either with high probability or if the common ratio is large enough, making them less attractive for low-access  computation. 

\item Other results in the paper include the following. 
\begin{itemize}
\item We demonstrate the possibility of reducing access further by jointly computing multiple two-valued linear combinations. We establish a connection between joint computation and a newly defined coding theoretic notion called \textit{generalized covering radius}~\cite{elimelech2021generalized}. 


\item A lower bound on the access requirement of quantized linear computations is derived, under the assumption that the user combines the downloaded entries linearly.

\item It is shown that the techniques developed in this paper for quantized linear computations can be adapted to the evaluation of multivariate monomials with powers restricted to a finite set. 
\end{itemize} 
\end{itemize}
 \subsection*{Organization of the Paper}

In Section~\ref{section:preliminaries}, we formally define the access-redundancy tradeoff problem for quantized linear computations. A brief background on covering codes and a summary of related works are also provided. Section~\ref{section:two_value} deals with two-valued linear computations. In Section~\ref{section:general}, we introduce the additive combinatorial notion of complexity and establish its connection with quantized linear computations. Section~\ref{section:complexity} explores this complexity notion in more detail. Section~\ref{section:generalized_covering} discusses the possibility of jointly obtaining multiple two-valued computations and its connection to generalized covering radius.  
 In Section~\ref{sec:monomials} we discuss the access-redundancy tradeoff for the evaluation of multivariate monomials with powers restricted to a finite set of values. 

 \section{Preliminaries}\label{section:preliminaries}
 This section begins with a formal problem statement and proceeds with an introduction to covering codes.  
 Since the subsequent schemes in the paper require covering codes with closure properties that have not been studied in the past, we construct several new families of covering codes using known tools\footnote{The novelty of these codes is limited to their closure properties, and they do not improve upon the best-known parameters.}, and hence a brief introduction to these tools is given as well. The last part of the section provides a summary of previous works related to the problem under consideration. 

We use standard notation for addition of sets, where for~$\cA,\cB\subseteq\bR$ we have~$\cA+\cB\triangleq\{a+b\vert a\in \cA,b\in \cB\}$, and~$\smallplus_{i}\cA_i$ is used for the addition of multiple sets~$\cA_i$. We also employ the notation $s\cA=\{sa\vert a\in \cA\}$.  Bold letters are used for vectors. Throughout this paper, vectors are row vectors, unless explicitly specified otherwise. Given a vector $\boldw \in \cA^k$, the entries of $\boldw$ are denoted by $(w_1, \dots, w_k)$.  
 We use the notation $\operatorname{supp}(\boldw) = \{i \in [k] \vert w_i \ne 0\}$, and use~$\1$ for the all-ones row vector.

 \subsection{Problem statement}\label{section:problemstatement}
	Given~$\boldx\in\bR^k$, we wish to encode it as~$\boldy\in\bR^n$ for~$n\ge k$, and store it in a distributed manner over~$n$ servers, one $\bR$-element~$y_i$ in each server\footnote{Typically, one would like to store multiple~$\boldx$'s, e.g., datapoints~$\boldx_1,\ldots,\boldx_N$ in a large dataset. In this paper, we focus on~$N=1$, and a general~$N$ follows by encoding all~$\boldx_i$'s in the same fashion.}. We focus on linear codes, in which every~$y_i$ is a linear combination of entries of~$\boldx$. 
For a given family of computations~$\cF$, we wish to devise an encoding mechanism such that any~$f\in\cF$ can be computed by accessing some~$\ell$ servers, where~$\ell$ is the \textit{access parameter} of the system.  While~$\cF$ is known at the time of encoding, the specific~$f\in\cF$ is not. Given the desired~$f\in\cF$ to compute, the user accesses~$\ell$ servers, whose identity is uniquely determined by~$f\in\cF$, downloads their content, and computes~$f(\boldx)$. The main goal in this field of study is to understand the resultant access-redundancy tradeoff by optimizing the correspondence between $n,k$ and~$\ell$ for families~$\cF$ of interest. Clearly, one would wish to minimize both~$n~$ and~$\ell$ simultaneously, but a clear tradeoff exists. In one end of the tradeoff we have~$n=|\cF|$ and~$\ell=1$ by pre-computing every~$f\in\cF$; in the other, we have~$n=\ell=k$, where the user downloads~$\boldx$ in its entirety and computes~$f(\boldx)$ locally.


In this paper, we focus on families of quantized linear computations over the real numbers, motivated by the rising prevalence of those in robust, efficient, and private machine learning inference (see \cite{galloway2017attacking,jia2020efficient,raviv2022information,raviv2021enhancing} and references therein). The access-redundancy tradeoff of quantized linear computations has been studied in the past~\cite{ho1998partial}, where the coefficients are restricted to $\{0,1\}$. The current paper looks at the general case where the coefficients can come from any finite set of real numbers. For finite set $\cA \subset \bR$, called the coefficient set, we define ~$$\cF_{\cA}=\{f(\boldx)=\boldw\boldx^\intercal\vert \boldw \in \cA^k\}.$$ 

For a given set of coefficients~$\cA$, a solution to this problem comes in the form of an encoding (i.e., storage) scheme, coupled with an algorithm that is given~$f\in\cF_\cA$,  outputs the identities of~$\ell$ servers which need to be contacted, alongside a recipe for combining their responses. We collectively refer to these operations, namely, the storage, access, and decoding, as an $\cA$-\textit{protocol} (protocol, in short). In order to understand the correspondence between~$n,k$, and~$\ell$ for a given~$\cA$, we consider the access ratio~$\ell/k$ and the redundancy ratio~$n/k$ in the asymptotic regime. 
\begin{definition}
For~$\alpha,\beta\in\bR$, we say that the pair~$(\alpha,\beta)$ is~$\cA$-feasible if there exists an infinite family of $\cA$-protocols with parameters~$\{ (k_i,n_i,\ell_i) \}_{i\ge 0}$ (with~$k_i$'s strictly increasing) such that~$\lim_{i\to \infty}(n_i/k_i,\ell_i/k_i)=(\alpha,\beta)$. 
\end{definition}
The goal of this paper is to understand the Pareto front of all~$\cA$-feasible pairs, for any given coefficient set~$\cA$. We will refer to a protocol as a \textit{universal protocol} 
if the encoding scheme is decided in advance without any knowledge about the quantization process, and it can support all possible coefficient sets but with potentially different access ratios. In Section~\ref{section:two_value}, we present a protocol for~$\cF_2\triangleq \cup_{a,b\in\bR}\cF_{\{a,b\}}$, i.e., a protocol which is only partially universal, where the same encoding  can accommodate all two-valued computations. This protocol relies on the existence of binary covering codes of short length with certain closure properties. Then we introduce a new additive combinatorial notion (referred to as \textit{complexity}) in Section~\ref{section:general}, which enables us to use the protocol for two-valued computations as a building block for a universal protocol. The next subsection presents the necessary background on covering codes to understand the two-valued protocol.

	\subsection{Covering codes}\label{section:coverigncodes}
	The covering radius (abbrv. radius) of a code is the minimum integer~$r$ such that Hamming balls of radius~$r$ that are centered in all codewords cover the entire space. Formally, for a code~$\cC$ of length~$p$ over an alphabet~$\Sigma$, the covering radius~$r=r(\cC)$ is defined as
	\begin{align*}
		r(\cC)=\min\{ r'\vert \cup_{\boldc\in\cC}B_H(\boldc,r')=\Sigma^p\}
	\end{align*}
	where~$B_H(\boldc,r')$ is the Hamming ball centered at~$\boldc$ with radius~$r'$. The covering radius is a fundamental topic in the study of error-correcting codes, and many constructions and bounds are well-known~\cite{cohen1997covering}.
	We briefly review some known techniques for constructing covering codes; for full details, the reader is referred to the respective references.
	
	The challenge in covering codes is constructing small codes with small radius. Clearly, the Cartesian product~$\cC_1\times \cC_2=\{ (\boldc_1,\boldc_2) \vert\boldc_1\in\cC_1,\boldc_2\in\cC_2\}$ of a code~$\cC_1$ with radius~$r_1$ and length~$p_1$ and a code~$\cC_2$ with radius~$r_2$ and length~$p_2$ has radius~$r_1+r_2$ and length~$p_1+p_2$. The following framework, called \textit{the amalgamated direct sum} and developed in~\cite{lobstein1989normal,graham1985covering,cohen1986further}, shows that under certain conditions, a code of radius~$r_1+r_2$ and length~$p_1+p_2-1$ can be constructed. This framework, described next, improves the rate of the Cartesian product (i.e., reduces its size-to-length ratio) without altering its radius, and yielded some of the best-known covering codes. 
 
    The norm~$N^{(i)}$ of entry~$i$ of a code~$\cC$ with radius~$r$ over~$\Sigma$ is defined as 
	\begin{align*}
		N^{(i)}\triangleq \max_{\boldw\in \Sigma^n}\left\{ \sum_{a\in\Sigma}d_H(\boldw,\cC_a^{(i)}) \right\},
	\end{align*}
	where~$\cC_a^{(i)}$ is the subset of~$\cC$ containing all codewords whose~$i$'th entry is~$a$, and~$d_H(\boldw,\cC_a^{(i)})\triangleq\min_{\boldc\in\cC_a^{(i)}}d_H(\boldw,\boldc)$, where~$d_H$ denotes the Hamming distance. If for some~$i$, we have~$N^{(i)}\le (r+1)|\Sigma|-1$, then the~$i$'th coordinate of~$\cC$ is called \textit{acceptable}, and~$\cC$ is called \textit{normal}. These technical tools play a role in the simple proof of the following theorem.
	
	\begin{theorem}[\cite{lobstein1989normal,graham1985covering,cohen1986further}]
		Let~$\cA,\cB$ be codes of length~$p_A,p_B$ and radii~$r_A,r_B$ over an alphabet~$\Sigma$. Further, the last coordinate of~$\cA$ and the first coordinate of~$\cB$ are acceptable, and~$\cA_{a}^{(p_A)},\cB_{a}^{(1)}\ne \varnothing$ for all~$a\in\Sigma$. Then, the amalgamated direct sum
		\begin{align*}
			\cA\dotplus\cB\triangleq \bigcup _{a\in\Sigma}\{ (\boldv,a,\boldw)\vert (\boldv,a)\in\cA,(a,\boldw)\in \cB \}
		\end{align*}
		is a code of length~$p_A+p_B-1$ and radius at most~$r_A+r_B$.
	\end{theorem}
	
	Suppose the codes~$\cA,\cB$ are linear codes with generator matrices~$G_A=[P \mid I],G_B=[I \mid P']$, respectively. The generator matrix of~$\cA\dotplus\cB$ is given by constructing a matrix which contains~$G_A$ and~$G_B$ on its diagonal (in block form), and $G_A$ and~$G_B$ intersect on one element, i.e., the lower-right element of~$G_A$ coincides with the upper-left element of~$G_B$. For example, if $$G_A=\begin{bmatrix}
	    1 & 1 & 0 & 0 \\
     	    1 & 0 & 1 & 0 \\
               	    1 & 0 & 0 & 1 
	\end{bmatrix}~~\text{and}~~G_B=\begin{bmatrix}
	    1 & 0 & 1  \\
     0 & 1 & 1
	\end{bmatrix},$$
 then generator matrix of $\cA\dotplus\cB$ takes the form: 
 $$\begin{bmatrix}
	    1 & 1 & 0 & 0 & 0 & 0\\
     	    1 & 0 & 1 & 0 & 0 & 0\\
               	    1 & 0 & 0 & 1 & 0 & 1 \\
                        0 & 0 & 0 & 0 & 1 & 1 \\
	\end{bmatrix}.$$
 
	
	Good covering radii are also obtained in \textit{piecewise constant codes}~\cite[Sec.~III.A]{cohen1986further}. In these codes the length~$n$ is partitioned as~$p=p_1+\ldots+p_t$, $p_i>0$ for all~$i$, and each codeword is partitioned respectively~$\boldc=(\boldc_1,\ldots,\boldc_t)$. A code  is piecewise constant if whenever it contains a word with
	\begin{align}\label{equation:partition}
		w_H(\boldc_1)=w_1,\ldots,w_H(\boldc_t)=w_t,
	\end{align}
	for some nonnegative integers~$w_1,\ldots,w_t$, where~$w_H$ denotes the Hamming weight,
	then it also contains \textit{all} such words. Constructing piecewise constant covering codes can be seen as covering a multi-dimensional array with Manhattan balls, as follows. Consider a~$t$-dimensional array in which the~$i$'th axis is indexed by~$0,1,\ldots,p_i$, and the~$(w_1,\ldots,w_t)$ entry contains the number of words satisfying~\eqref{equation:partition}, i.e.,~$\prod_{j=1}^{t}\binom{p_j}{w_j}$. It is an easy exercise to show that if one manages to cover this array using Manhattan balls of radius~$r$ centered at some~$m$ entries~$\{ (w_{i,1},\ldots,w_{i,t}) \}_{i=1}^m$, then the union of the~$m$ sets of words corresponding to these~$m$ entries is a piecewise constant code of radius~$r$. An example of such construction is given in Fig.~\ref{figure:twoDimArr} in Section~\ref{section:two_value} which follows.

	\subsection{Related work}
	Variants of the access-redundancy tradeoff problem have been studied in many previous works. First, the~$\bF_q$ variant of the problem, where entries of~$\boldx$ is over some finite field~$\bF_q$ and~$\cF=\{ f(\boldx)=\boldw\boldx^\intercal\vert \boldw\in\bF_q^k \}$ is a folklore result in the area. It can be shown~\cite{cohen1997covering} that a protocol with a given~$n,k,\ell$ as above exists if and only if there exists a linear code of length~$n$, dimension~$n-k$, and covering radius~$r=\ell$. However, this method cannot be applied for linear computation over real numbers, which is the focus of the current paper. In the context of polynomial evaluation,  the following question was asked in \cite{KedlayaU11}:  a given
polynomial $f$ is to be stored so that any evaluation $f(x)$ can be computed with low access. The result  addresses a single, yet very attractive, point of the access-redundancy tradeoff for single polynomial evaluation.
Some access problems were asked in the context of  polynomial computations~\cite{yu2019lagrange}, where~\cite{raviv2019download} offers a decoding technique for reducing download in certain cases.

	When it comes to linear computation over~$\bR$, the family~$\cF_{\{0,1\}}=\{ f(\boldx)=\boldw\boldx^\intercal \vert \boldw\in\{ 0,1 \}^k\}$ has been studied in~\cite{ho1998partial}, and covering codes are employed in a different way than the finite field case. 
	In Section~\ref{section:equivalence}, it is shown that in fact, all two-valued linear computations are equivalent. Furthermore, our results strictly improve (i.e., both in terms of redundancy and in terms of access) all the results from~\cite{ho1998partial}. It should be noted that even though~\cite{ho1998partial} uses the term ``partial sum'' to describe $\cF_{\{0,1\}}$, an identical term is widely used~\cite{fredman1982complexity,patrascu2004tight} for a different problem where~$\cF=\{ f_{i,j}(\boldx)=\sum_{\ell=i}^jx_i\vert 1\le i\le j\le n \}$ (i.e., consecutive indices, which in some variations only begin with~$1$), or its multidimensional variant~\cite{chazelle1989computing}. The access aspect of these computations is often studied under the name  ``cell-probe,'' e.g.,~\cite{puatracscu2010cell}. The cell probe model \cite{Yao81a} is a computational framework where all operations  are free, except memory cell access.  In this model, data is pre-processed into a data structure, and computation is equivalent to querying a set of memory cells.  This approach has resulted in interesting computational complexity lower bounds for a wide variety of problems including those in \cite{Ajtai88, BeameF02, PatrascuD06,  puatracscu2010cell}. 

	More broadly, our results are also related to recent trends in \textit{coded computation}. This recently popularized area \cite{lee2017speeding, yu2019lagrange} studies various distributed computation tasks under the \textit{straggler} effect---the user accesses \textit{all} servers, and completes the computation from the responses of the fastest ones. This work can be seen as a complementary study that addresses the \textit{number} of servers that should be accessed. In the straggler problem,  the identity of the fastest servers is
		not known a priori, whereas, in the access problem under consideration here, the identity of the servers that are accessed is a deterministic function of the computation task, and these servers are not necessarily faster than others.
	Extending the results in this paper to address the straggler effect, is left for future work.


 Coding techniques to reduce access requirements have previously been investigated in other settings as well. Locally recoverable codes \cite{GopalanHSY12, TamoB14}  are codes designed to recover the contents of a failed node in a distributed storage system by accessing a small number of  surviving nodes. Locally recoverable codes can be easily described in the computation framework we discuss in this paper.  Constructing these codes is essentially equivalent to finding a generator matrix $G$ with columns $\boldg_1,\dots,\boldg_n$ such that the function family $\cF = \{f(x) = g_ix^T \mid i \in [n] \}$ has a low-access and straggler-resilient protocol. Locally decodable codes \cite{Yekhanin12} are codes for which every message symbol can be decoded by accessing a small fraction of code symbols of a possibly corrupted codeword.  Specifying to the function family $\cF = \{f(x) = x_i \mid i \in [k]\}$ under an adversarial setting yields the problem of locally decodable codes.
Compression with locality  \cite{MazumdarCW15} is an information-theoretic notion that focuses on source coding a collection of random variables such that each individual  value can be decoded with low access.

	\section{Two-Valued Linear Computations}   \label{section:two_value}

 In this section, we focus on quantized linear computations over~$\bR$, whose coefficients are quantized to two different values (e.g.,~$\{\pm1\}$). All such linear computations which only require two real coefficients are equivalent in a sense that will be clarified shortly in Section~\ref{section:equivalence}. In Section~\ref{section:pm1}, we provide a general framework for constructing $\{\pm1\}$ protocols based on covering codes with certain closure properties. We present various such covering codes and the resulting feasible pairs. From a practical perspective, it is preferable that the encoding is systematic, i.e., for each $x_i$ there exists a node storing it in the uncoded form. Quantized linear computations may not be the only computation performed on the dataset, and systematic storage ensures the availability of raw data to perform other computations as well. The protocols in \cite{ho1998partial} and the protocols we present in Section~\ref{section:pm1} (that outperform the results in \cite{ho1998partial}) employ systematic storage. However, if non-systematic encoding schemes are permitted, it is possible to do better than our Pareto optimal front under the systematic approach, as demonstrated in Section~\ref{sec:non_sys}. In Section~\ref{section:bound}, we provide lower bounds under the assumption that the user combines the~$\ell$ downloaded entries of~$\boldy$ linearly. Section~\ref{sec:convex} shows that points on the line joining two feasible access-redundancy pairs are also feasible.

	\subsection{Two-value equivalence}\label{section:equivalence}

	
	\begin{proposition}\label{proposition:2equivalence}
		For any distinct~$a,b\in \bR$, a pair~$(\alpha,\beta)$ is $\{\pm1\}$-feasible if and only if it is $\{ a,b \}$-feasible. Furthermore, data stored using an $\cF_{\{\pm 1\}}$-protocol can be used to retrieve~$\boldw\boldx^\intercal$ for any~$\boldw\in\{a,b\}^k$ and any distinct~$a,b\in\bR$ by using at most one additional node.
	\end{proposition}
	
		\begin{proof}
	Given an $\cF_{a,b}$-protocol, store~$\boldx$ according to it, with an additional node containing~$\1 \boldx^\intercal$ (if not already present). Then, given any~$\boldw\in\{\pm1 \}^k$, let~$\boldw'\in \{a,b\}^k$ be such that~$w'_i=a$ if~$w_i=1$, and~$w'_i=b$ if~$w_i=-1$. By following the protocol for retrieving~$\boldw'\boldx^\intercal$, and retrieving~$\1 \boldx^\intercal$ from its designated node, the user can compute~
	\begin{eqnarray*}
		&&	\tfrac{2}{a-b}\cdot\boldw'\boldx^\intercal-\tfrac{a+b}{a-b}\cdot\1\boldx^\intercal\\ &=&
		\sum_{j\vert w_j=1}\tfrac{2a}{a-b}x_j+\sum_{j\vert w_j=-1}\tfrac{2b}{a-b}x_j-\tfrac{a+b}{a-b}\1\boldx^\intercal\\	
		&=& \sum_{j\vert w_j=1}\tfrac{2a-a-b}{a-b}x_j+\sum_{j\vert w_j=-1}\tfrac{2b-a-b}{a-b}x_j=\boldw\boldx^\intercal.
	\end{eqnarray*}
	
	Conversely, given an $\cF_{\{\pm 1\}}$-protocol, store~$\boldx$ according to it, with an additional node containing~$\1 \boldx^\intercal$ (if not already present). Then, given any~$\boldw\in\{a,b \}^k$, let~$\boldw'\in\{\pm1\}^k$ such that~$w'_i=1$ if~$w_i=a$, and $w'_i=-1$ if~$w_i=b$. By following the protocol for retrieving~$\boldw'\boldx^\intercal$, and retrieving~$\1\boldx^\intercal$ from its designated node, the user can compute
	\begin{align}\label{equation:+-equiv}
		&\tfrac{a-b}{2}\boldw'\boldx^\intercal+\tfrac{a+b}{2}\1\boldx^\intercal\nonumber\\
		&=\tfrac{a-b}{2}\sum_{j\vert w_j=a}x_i-\tfrac{a-b}{2}\sum_{j\vert w_j=b}x_j+\tfrac{a+b}{2}\1\boldx^\intercal\nonumber\\
		&=\sum_{j\vert w_j=a}x_i\left( \tfrac{a-b}{2}+\tfrac{a+b}{2} \right)+\sum_{j\vert w_j=b}x_i(\tfrac{b-a}{2}+\tfrac{b+a}{2})=\boldw\boldx^\intercal.
	\end{align}
	
	
	In both directions, since only at most one additional node is used for storage, and at most one additional node is accessed during computation, it follows that the protocols give rise to identical feasible pairs due to the limit operation, and the claim follows. The ``furthermore'' part is clear from~\eqref{equation:+-equiv}.
\end{proof}
	
	By transitivity, Proposition~\ref{proposition:2equivalence} readily implies that all two-valued computations have the same set of feasible pairs, hence the same Pareto front in the access-redundancy regime. Moreover, one can devise a protocol exclusively for~$\cF_{\{\pm 1\}}$ and then use it for~$\cF_{\{a,b\}}$ for any distinct~$a,b\in \bR$. Therefore, an~$\cF_{\{\pm 1\}}$-protocol can be easily extended to an~$\cF_2$-protocol, where 
	$\cF_2\triangleq\cup_{a\ne b}\cF_{\{a,b\}}$. In the sequel, we focus on~$\cF_{\{\pm 1\}}$ due to favorable closure properties (described shortly), which by the above discussion extend to $\cF_2$. In particular, our techniques produce families of codes that not only improve upon existing ones (that were tailored exclusively for~$\cF_{\{0,1\}}$~\cite{ho1998partial}) both in the access ratio and the redundancy ratio but also extend them to all two-valued computations.
	
	\subsection{Systematic storage  approach}\label{section:pm1}
	As a motivating example (noted in~\cite{ho1998partial}), it is readily seen that the pair~$(1,0.5)$ is~$\{0,1\}$-feasible (and hence $\cF_2$-feasible) by simply storing an additional node containing~$\1\boldx^\intercal$, on top of a node for each~$x_i$ (i.e., $n=k+1$). Then, if~$w_H(\boldw)\le k/2$, one can compute~$\boldw\boldx^\intercal$ by accessing at most~$k/2$ entries from the systematic part. Otherwise, it is answered by accessing the node containing $\1\boldx^\intercal$, all the nodes~$x_i$ such that~$w_i=0$, and subtracting; a respective protocol for any two-value computation then follows from Proposition~\ref{proposition:2equivalence}.  This example can be seen as a special case of the following framework,
	which provides a low-access protocol from any binary covering code, and is particularly useful when the underlying covering code is closed under complement. Intuitively, when employing binary covering codes in their~$\{\pm1\}$-representation, closure under complement translates to negation over the reals, and approximately half of the storage costs can be saved; this is the crux of the improvement over~\cite{ho1998partial}. In the sequel several suitable covering code constructions are given, alongside a comparison to~\cite{ho1998partial}.

	As mentioned earlier, we focus on protocols for~$\cF_{\{\pm 1\}}$, which by Proposition~\ref{proposition:2equivalence} imply protocols for all two-valued computations. To this end, we use~$\bF_2$-arithmetic, and refer to vectors over~$\bF_2$ in their~$\{\pm 1\}$-representation, i.e., use the real~$1$ instead of the Boolean~$0$ and the real~$-1$ instead of the Boolean~$1$. 
	The framework is based on the following simple definition, in which~$\oplus$ denotes addition in~$\bF_2$ (i.e., point-wise exclusive OR).
	
	\begin{definition}\label{definition:complementFree}
		For a code~$\cC$ over~$\bF_2$, let~$\hat{\cC}\subseteq \cC$ which contains exactly one of~$\{ \boldc,\boldc\oplus \1 \}$, for every~$\boldc\in \cC$. That is, if~$\{ \boldc,\boldc\oplus\1 \}\subseteq \cC$, then exactly one of~$\{\boldc,\boldc\oplus\1\}$ is in~$\hat{\cC}$. If~$\boldc\oplus\1\notin \cC$ for some~$\boldc\in\cC$, then~$\boldc$ is also in~$\hat{\cC}$. When the code~$\cC$ is clear from the context, we denote~$\hat{c}\triangleq|\hat{\cC}|$.	
	\end{definition}
	Clearly, there are multiple ways to generate~$\hat{\cC}$ from~$\cC$, all of which result in~$\hat{\cC}$ of the same size. Since only the size of~$\hat{\cC}$ matters in our context, we assume some unspecified canonical way of constructing a unique~$\hat{\cC}$ from~$\cC$. As an example, consider the following observation.
	
	\begin{observation} \label{observation:complement}
		A (not-necessarily linear) code~$\cC$ that is closed under complement (i.e.,~$\boldc\in\cC$ if and only if~$\1\oplus\boldc\in\cC$) satisfies~$|\hat{\cC}|=|\cC|/2$. Notice that a linear code~$\cC$ is closed under complement if and only if~$\1\in\cC$.
	\end{observation}
		
	This gives rise to the following theorem.
	
	\begin{theorem}\label{theorem:OLAPimprov}
		If there exists a (not-necessarily linear) code~$\cC$ of length~$p$ and covering radius~$r$ over~$\bF_2$, then the pair~$(\frac{p+\hat{c}}{p},\frac{r+1}{p})$ is $\{\pm1\}$-feasible. Moreover, this pair can be obtained using systematic storage schemes.
	\end{theorem}
	
	\begin{proof}
	Construct the matrix~$M=(B\vert I)\in \{0,\pm 1\}^{p\times (p+\hat{c})}$, where~$I$ is the identity matrix and~$B$ contains all~$\hat{c}$ vectors of~$\hat{\cC}$ in their~$\{\pm 1\}$-representation as columns. For~$t\in\bN$ consider~$k=tp$, and partition~$\boldx\in\bR^{tp}$ to~$t$ parts~$\boldx_1,\ldots,\boldx_{t}$ of size~$p$ each. To encode, let~$\boldy_i\triangleq \boldx_i M$, and let~$\boldy=(\boldy_1,\ldots\boldy_t)\in \bR^{t(p+\hat{c})}$. It is easy to see that this encoding is systematic. The storage overhead of the resulting scheme is clearly~$n/k=\frac{p+\hat{c}}{p}$. 
	
	To retrieve a product~$\boldw\boldx^\intercal$ for some~$\boldw\in\{ \pm 1 \}^k$, write~$\boldw=(\boldw_1,\ldots,\boldw_t)$ with~$\boldw_i\in\{ \pm 1 \}^p$ for all~$i$ and note that $\boldw\boldx^\intercal=\sum_{i=1}^{t}\boldw_i\boldx_i^\intercal$.
  Each~$\boldw_i\boldx_i^\intercal$ can be retrieved by accessing at most~$r+1$ entries from~$\boldy_i$, as follows. Since~$\cC$ is a covering code of radius~$r$, there exists~$\boldc_i\in\cC$ such that~$d_H(\boldc_i,\boldw_i)\le r$. Consider two cases:
	\begin{itemize}
		\item [(a)] If $\boldc_i\in \hat{\cC}$, the user accesses the node which stores~$\boldc_i\boldx_i^\intercal$, and at most~$r$ which store~$x_{i,j}$ for indices~$j\in[p]$ on which~$\boldc_i$ and~$\boldw_i$ differ. The user computes
		\begin{align*}
			&\boldc_i\boldx_i^\intercal+2\sum_{j\vert c_{i,j}=- w_{i,j}}w_{i,j}x_{i,j}=\\
			&\sum_{j\vert c_{i,j}=w_{i,j}}w_{i,j}x_{i,j}-\sum_{j\vert c_{i,j}=-w_{i,j}}w_{i,j}x_{i,j}+\\
			&2\cdot\sum_{j\vert c_{i,j}=- w_{i,j}}w_{i,j}x_{i,j}=\boldw_i\boldx_i^\intercal.
		\end{align*}
		\item [(b)] If~$\boldc_i\notin\hat{\cC}$ it follows that the vector~$\boldc_i'\triangleq\boldc_i\oplus \1$ is in~$\hat{\cC}$ (see Definition~\ref{definition:complementFree}). The user accesses the node which stores~$\boldc_i'\boldx_i^\intercal$, and at most~$r$ nodes which store~$x_{i,j}$ for indices~$j\in[p]$ on which~$\boldc_i$ and~$\boldw_i$ differ. The user computes
		\begin{align*}
			&-\boldc_i'\boldx_i^\intercal+2\sum_{j\vert c_{i,j}=- w_{i,j}}w_{i,j}x_{i,j}\\
			&= (\boldc_i'\oplus \1)\boldx_i^\intercal+2\sum_{j\vert c_{i,j}=- w_{i,j}}w_{i,j}x_{i,j} \\
			&=	\boldc_i\boldx_i^\intercal+2\sum_{j\vert c_{i,j}=- w_{i,j}}w_{i,j}x_{i,j} =\boldw_i\boldx_i^\intercal.
		\end{align*}
	\end{itemize}
	Clearly, in both cases, the user accesses at most~$r+1$ nodes, hence~$\ell/k=(r+1)t/pt$, which concludes the proof.
\end{proof}

\begin{figure*}[ht!]
		\begin{subfigure}{0.5\textwidth}
			\centering
			\resizebox{.85 \textwidth}{!} 
			{
				\begin{tabular}{|c|c|c|}
					\hline
					\rowcolor[HTML]{C0C0C0} 
					\textbf{Redundancy~$n/k$} & \textbf{Access~$\ell/k$} & \textbf{Code name}         \\ \hline
					1                               & 0.5                              & Trivial                    \\ \hline
					1.17                         & 0.48                            & $\texttt{PiecewiseAmal}_9$ \\ \hline
					1.21                         & 0.47                             & $\texttt{PiecewiseAmal}_7$ \\ \hline
					1.25                         & 0.46                             & $\texttt{NonlinAmal}_9$    \\ \hline
					1.27                         & 0.45                             & $\texttt{NonlinAmal}_8$    \\ \hline
					1.32                         & 0.44                             & $\texttt{HamAmal}_9$       \\ \hline
					1.35                         & 0.43                             & $\texttt{HamAmal}_8$       \\ \hline
					1.42                         & 0.42                             & $\texttt{HamAmal}_6$       \\ \hline
					1.47                         & 0.41                             & $\texttt{HamAmal}_5$       \\ \hline
					1.53                         & 0.4                             & $\texttt{HamAmal}_4$       \\ \hline
					1.62                         & 0.38                            & $\texttt{HamAmal}_3$       \\ \hline
					1.73                         & 0.36                             & $\texttt{HamAmal}_2$       \\ \hline
					1.89                         & 0.33                             & $\texttt{HamAmal}_1$       \\ \hline
					2.14                         & 0.29                             & $\texttt{HamExp}_0$        \\ \hline
					3.0                          & 0.25                             & $\texttt{HamExp}_1$        \\ \hline
					4.2                        & 0.2                             & $\texttt{HalfSpace}_5$        \\ \hline
					6.33                         & 0.17                              & $\texttt{HalfSpace}_6$        \\ \hline
				\end{tabular}
			}
			\caption{}\label{table:Pareto}
		\end{subfigure}
		\begin{subfigure}{0.5\textwidth}
			\centering
	\includegraphics[scale=0.55]{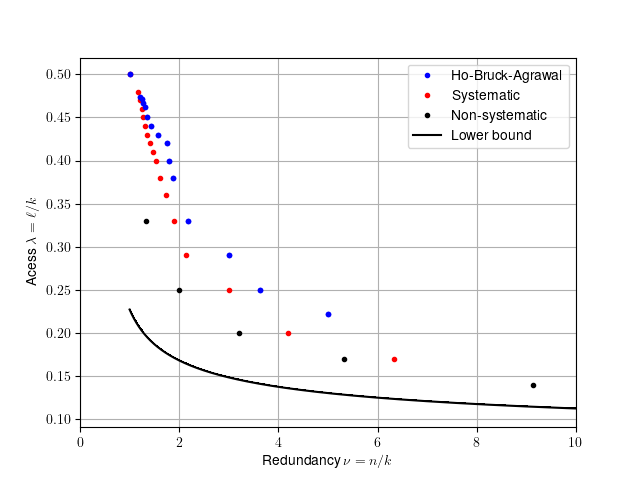}
			\caption{}\label{figure:ShukiComparison}
		\end{subfigure}\caption{(a) The Pareto optimal front of the suggested systematic solutions in Section~\ref{section:Hamming}, Section~\ref{section:HalfSpace}, Section~\ref{section:knownCov}, and Section~\ref{section:newCode},  truncated to two decimal points, when varying~$i$ from~$0$ to~$9$ and omitting solutions whose redundancy factor is not feasible ($\nu=n/k>10$). (b) Graphical depiction of our solutions, and comparison to~\cite{ho1998partial}. In particular, the red points are the Pareto optimal front of all systematic solutions and the blue points are the results from~\cite[Table~9]{ho1998partial}. It can be seen that all blue points, except the trivial point, are dominated by some red points. The black points are the non-systematic solutions in Section~\ref{sec:non_sys}. The black line is a numeric evaluation of the bound in~\eqref{equation:boundFinal}. 
		}\label{figure:ShukiComparisonFig}
	\end{figure*}

	In the remaining part of this subsection, we present corollaries of the above theorem using linear and nonlinear covering codes. The resulting Pareto optimal front and a comparison to~\cite{ho1998partial} are summarized in Fig.~\ref{figure:ShukiComparisonFig}. 
	\subsubsection{Hamming codes}\label{section:Hamming}
	It is well-known that the~$[7,4]_2$ Hamming code is a perfect code of minimum distance~$3$, and hence also a covering code of radius~$1$. Furthermore, by the amalgamated construction of~\cite[Thm.~20.i]{graham1985covering}, the~$[7,4]_2$ Hamming code can be extended any number~$i\ge 0$ of times to a linear code of the same dimension~$4$, length~$7+2i$ and covering radius~$1+i$. This is done using an amalgamated direct sum (see Section~\ref{section:coverigncodes} for definition) with the repetition code of length~$2i+1$ (whose covering radius is~$i$); more specifically, by extending the bottom row of the generator matrix by two~$1$'s at a time, and extending the top three rows by two~$0$'s at a time:
 \begin{align*}
			\begin{pmatrix}
				1&1&0&1&0&0&0\\
				1&0&1&0&1&0&0\\
				0&1&1&0&0&1&0\\
				1&1&1&0&0&0&1\\
			\end{pmatrix}\mapsto
			\begin{pmatrix}
				1&1&0&1&0&0&0&0^{2i}\\
				1&0&1&0&1&0&0&0^{2i}\\
				0&1&1&0&0&1&0&0^{2i}\\
				1&1&1&0&0&0&1&1^{2i}\\
			\end{pmatrix}.
	\end{align*}
	The resulting extended codes are linear by definition, and clearly contain~$\1$, and hence are closed under complement (see Observation~\ref{observation:complement}). Therefore, they can be used in Theorem~\ref{theorem:OLAPimprov} with~$\hat{c}=|\cC|/2=8$, $r=1+i$, and~$p=7+2i$. The resulting feasible pairs are $\{ (\frac{15+2i}{7+2i},\frac{2+i}{7+2i}) \}_{i\ge 0}$, the~$i$'th of which is referred to as~$\texttt{HamAmal}_i$.
	
	Additionally, for any covering code~$\cC$ of length~$p$, covering radius~$r$, and size~$s$, the code~$\cD_i\triangleq\cC\times \bF_2^i$ is a covering code of length~$p+i$, size~$2^is$, and identical covering radius~$r$. This follows easily by using a (not-amalgamated) direct sum of covering codes, and the fact that the covering radius of~$\bF_2^i$ is zero. 
	Moreover, it is an easy exercise to show that~$\hat{d}=\hat{c}2^i$. Applying this method with~$\cC$ being the~$[7,4]_2$ Hamming code results in covering codes~$\cD_i$ of length~$7+i$ and covering radius~$1$, for which~$\hat{d}_i=2^{i+3}$. In turn, the codes~$\{\cD_i\}_{i\ge 0}$ give rise to the feasible pairs~$\{ (\frac{2^{i+3}+i+7}{i+7},\frac{2}{i+7}) \}_{i\ge 0}$ by Theorem~\ref{theorem:OLAPimprov}, the~$i$'th of which is referred to as~$\texttt{HamExp}_i$.
	
	\subsubsection{Half space}  \label{section:HalfSpace}
		The entire space~$\cC=\bF_2^p$ is a covering code of size~$2^p$ and covering radius~$r=0$. Clearly, $\cC$ is closed under complement and ~$\hat{c}=2^{p-1}$. Therefore, by Theorem~\ref{theorem:OLAPimprov} the pairs~$\{ (\frac{i+2^{i-1}}{i},\frac{1}{i}) \}_{i\ge 1}$ are feasible, the~$i$'th of which is referred to as~$\texttt{HalfSpace}_i$.
	
	\subsubsection{Known nonlinear covering codes} \label{section:knownCov}
	Ref.~\cite{cohen1986further} further extended the amalgamated construction technique of~\cite{graham1985covering} to nonlinear codes. Specifically, a certain $12$-word nonlinear code of length~$6$ and covering radius~$1$ can be extended any number~$i\ge 0$ of times (by amalgamating it with the repetition code of length~$2i+1$) to get a $12$-word code of length~$6+2i$ and covering radius~$1+i$,
	as shown below. 
			\begin{align*}
			\begin{matrix}
				\text{Word }1:&0&0&0&1&0&0&0^{2i}\\
				\text{Word }2:&0&0&0&0&1&0&0^{2i}\\
				\text{Word }3:&0&0&0&0&0&1&1^{2i}\\
				\text{Word }4:&1&0&0&1&1&1&1^{2i}\\
				\text{Word }5:&0&1&0&1&1&1&1^{2i}\\
				\text{Word }6:&0&0&1&1&1&1&1^{2i}\\
				\text{Word }7:&1&1&1&0&1&1&1^{2i}\\
				\text{Word }8:&1&1&1&1&0&1&1^{2i}\\
				\text{Word }9:&1&1&1&1&1&0&0^{2i}\\
				\text{Word }10:&0&1&1&0&0&0&0^{2i}\\
				\text{Word }11:&1&0&1&0&0&0&0^{2i}\\
				\text{Word }12:&1&1&0&0&0&0&0^{2i}\\
			\end{matrix}	
		\end{align*}
	It is readily verified that all the extensions are closed under complement; row~$j$ is the complement of row~$6+j$ for every~$j\in[6]$. Therefore, Theorem~\ref{theorem:OLAPimprov} can be used with~$\hat{c}=6$, $r=1+i$, and~$p=6+2i$ to obtain that the pairs~$\{ (\frac{12+2i}{6+2i},\frac{2+i}{6+2i}) \}_{i\ge 0}$ are feasible, the~$i$'th of which is referred to as~$\texttt{NonlinAmal}_i$.
	
	\subsubsection{New nonlinear covering codes}\label{section:newCode}
	\begin{figure}[ht!]
  \centering
		\renewcommand{\arraystretch}{1.5}
		\begin{tabular}{cccccc}
			&                          & \multicolumn{4}{c}{$w_2$}                                                                         \\
			&                          & $0$                    & $1$                    & $2$                    & $3$                    \\ \cline{3-6} 
			\multirow{3}{*}{$w_1$} & \multicolumn{1}{c|}{$0$} & \multicolumn{1}{c|}{$1$} & \multicolumn{1}{c|}{\cir{$3$}} & \multicolumn{1}{c|}{$3$} & \multicolumn{1}{c|}{\cir{$1$}} \\ \cline{3-6} 
			& \multicolumn{1}{c|}{$1$} & \multicolumn{1}{c|}{$2$} & \multicolumn{1}{c|}{$6$} & \multicolumn{1}{c|}{$6$} & \multicolumn{1}{c|}{$2$} \\ \cline{3-6} 
			& \multicolumn{1}{c|}{$2$} & \multicolumn{1}{c|}{\cir{$1$}} & \multicolumn{1}{c|}{$3$} & \multicolumn{1}{c|}{\cir{$3$}} & \multicolumn{1}{c|}{$1$} \\ \cline{3-6} 
		\end{tabular}\caption{The two-dimensional array used for the~$8$-word code of length~$5$ and covering radius~$1$ given in Section~\ref{section:newCode}. It is easily verified that Manhattan balls of radius~$1$ centered at the circled entries cover the entire table. These balls correspond to a covering code of radius~$1$ and size~$1+3+3+1=8$, see~\cite[Sec.~III]{cohen1986further} for the full details.}\label{figure:twoDimArr}
	\end{figure}
	We show that additional pairs are feasible by constructing a new covering code that is closed under complement, using the tools in~\cite{cohen1986further}. We then verify the code's normality
	with a simple computer program, and extend it using the amalgamation technique. Specifically, using the two-dimensional array in Fig.~\ref{figure:twoDimArr}, we build an $8$-word code of length~$5$ and covering radius~$1$, and then extend it any number~$i\ge 0$ of times to get an~$8$-word code of length~$5+2i$ and covering radius~$1+i$, as follows.
	\begin{align*}
		\scalebox{0.9}{$
			\begin{matrix}
				\text{Word }1:&0&0&1&0&0&0^{2i}\\
				\text{Word }2:&0&0&0&1&0&0^{2i}\\
				\text{Word }3:&0&0&0&0&1&1^{2i}\\
				\text{Word }4:&0&0&1&1&1&1^{2i}\\
				\text{Word }5:&1&1&0&1&1&1^{2i}\\
				\text{Word }6:&1&1&1&0&1&1^{2i}\\
				\text{Word }7:&1&1&1&1&0&0^{2i}\\
				\text{Word }8:&1&1&0&0&0&0^{2i}\\
			\end{matrix} $} 
	\end{align*}
	
	It is readily verified that all the extensions are closed under complement; row~$j$ is the complement of row~$4+j$ for every~$j\in[4]$. Therefore, Theorem~\ref{theorem:OLAPimprov} can be used with~$\hat{c}=4$, $r=1+i$, and~$p=5+2i$ to obtain that the pairs~$\{ (\frac{9+2i}{5+2i},\frac{2+i}{5+2i}) \}_{i\ge 0}$ are feasible, the~$i$'th of which is referred to as~$\texttt{PiecewiseAmal}_i$.

			\subsection{A non-systematic approach}\label{sec:non_sys}
			If the systematic storage requirement is relaxed, then it is possible to obtain feasible pairs that dominate most of the systematic approach  Pareto optimal pairs (see Fig.~\ref{figure:ShukiComparisonFig}).  The protocol corresponding to  Theorem~\ref{theorem:OLAPimprov} does not require access to systematic symbols if the covering radius of $\cC$ is zero. Therefore, if non-systematic encoding schemes are allowed, one can remove systematic nodes from $\texttt{HalfSpace}_i$ to obtain $\{\pm1\}$-feasible pairs~$\{(\frac{2^{i-1}}{i},\frac{1}{i})\}_{i \ge 1}$. These non-systematic solutions are summarized in Fig.~\ref{figure:ShukiComparisonFig}.

   	\subsection{A lower bound} \label{section:bound}
	It is well-known that the maximum number of~$\{\pm1\}$ vectors that belong to any given~$\ell$-dimensional $\bR$-subspace is~$2^\ell$ (e.g., \cite[Lemma~7]{raviv2022information}). Assuming that the user linearly combines the data that is downloaded from the nodes, this gives rise to the following bound. 
		
		\begin{theorem}\label{theorem:bound}
			An~$\cF_2$-protocol with a given~$n$, $k$,~$\ell$, and linear decoding must satisfy that~$\binom{n}{\ell}2^\ell\ge 2^k$.
		\end{theorem}
				\begin{proof}
			According to the discussion in Section~\ref{section:equivalence}, there exists a protocol for~$\cF_2$ if and only if there exists a protocol for~$\cF_{\{\pm 1\}}$ with identical parameters, and hence we may focus on protocols for~$\cF_{\{\pm 1\}}$. By the definition of a protocol (Section~\ref{section:problemstatement}), each of the~$2^k$ vectors~$\boldw\in\{\pm1\}^k$ has a corresponding set of~$\ell$ nodes which must be accessed to retrieve~$\boldw\boldx^\intercal$. Clearly, there are at most~$\binom{n}{\ell}$ such sets, each of which contains at most~$2^\ell$ $\{\pm1\}$-vectors in its span.
		\end{proof}
		
		We proceed to evaluate the above bound with respect to the constructions in Section~\ref{section:pm1}. Theorem~\ref{theorem:bound} implies that
		\begin{align}\label{equation:bound1}
			\frac{\ell}{k}+\frac{\log\binom{n}{\ell}}{k}\ge 1. 
		\end{align}
		Denote~$\nu\triangleq \frac{n}{k}$ and~$\lambda\triangleq\frac{\ell}{k}$; in light of the results in Section~\ref{section:pm1} we restrict our attention to~$1\le \nu\le 10$ and~$0<\lambda\le 1/2$ (see Fig.~\ref{figure:ShukiComparisonFig}). A known bound asserts that
		\begin{align*}
			\nu k H(\lambda/\nu)-\log(\nu k+1)\le\log\binom{\nu k}{\lambda k}\le \nu k H(\lambda/\nu),
		\end{align*}
		where~$H$ is the binary entropy function, and hence in the regime where~$k$ is large we may use the approximation
		\begin{align*}
			\frac{\log\binom{n}{\ell}}{k}=\frac{\log\binom{\nu k}{\lambda k}}{k}\approx \nu H(\lambda/\nu).
		\end{align*}
		Plugging this approximation into~\eqref{equation:bound1} implies that
		\begin{align}\label{equation:boundFinal}
			H(\lambda/\nu)\ge \frac{1-\lambda}{\nu}.
		\end{align}
		A numeric evaluation of this bound is given in Fig.~\ref{figure:ShukiComparisonFig}. It is apparent that even under the assumption of linear decoding a substantial gap exists between our constructions and bounds. Under systematicity assumption, we can improve the bound in Theorem~\ref{theorem:bound} to $\binom{n}{\ell}-\binom{k}{\ell} \ge 2^{k-\ell}$, however it results in negligible improvements when $k$ is large. 

	\subsection{Convex combinations}\label{sec:convex}
  We now show that points on the line connecting any two feasible pairs obtained using protocols with finite parameters are also feasible. 
 \begin{proposition}\label{prop:combination}
		Suppose there exist two $\{\pm1\}$-protocols $P1$ and $P2$ with finite parameters $(k_1,n_1,\ell_1)$ and $(k_2,n_2,\ell_2)$ respectively. Then for all positive integers $u,v$, the pair~$\Big(\frac{u n_1+vn_2}{u k_1+v k_2},\frac{u \ell_1+v \ell_2}{u k_1+v k_2}\Big)$ is $\{\pm1\}$-feasible.  
 \end{proposition}
\begin{proof}
	 For~$t\in\bN$ consider~$k=t(uk_1+vk_2)$, and partition~$\boldx\in\bR^{k}$ to~$t(u+v)$ parts~$\boldx_1,\ldots,\boldx_{t(u+v)}$ such that the first $tu$ parts are of size~$k_1$ and the remaining $tv$ parts are of size $k_2$.  For all $i \in [tu]$, encode $\boldx_i$ to $\boldy_i \in \bR^{n_1}$ using the encoding scheme of protocol $P1$. Similarly for $tu+1 \le i \le t(u+v)$, employ the encoding scheme of protocol $P2$ on $\boldx_i$ to get $\boldy_i\in \bR^{n_2}$. Hence, $\boldy=(\boldy_1,\dots,\boldy_{t(u+v)}) \in \bR^{un_1+vn_2}$.   
  The storage overhead of the resulting scheme is clearly~$$n/k=\frac{u n_1+vn_2}{u k_1+v k_2}.$$ To retrieve ~$\boldw\boldx^\intercal$ for some~$\boldw\in\{ \pm 1 \}^k$, write~$\boldw=(\boldw_1,\ldots,\boldw_{t(u+v)})$ with~$\boldw_i\in\{ \pm 1 \}^{k_1}$ for all~$i \in [tu]$ and $\boldw_i\in\{ \pm 1 \}^{k_2}$ for the rest.    
  For each~$i \in [tu]$, using protocol $P1$, ~$\boldw_i\boldx_i^\intercal$ can be computed by accessing at most~$\ell_1$ entries from~$\boldy_i$. Similarly, for $tu+1 \le i \le t(u+v)$, ~$\boldw_i\boldx_i^\intercal$ can be retrieved by accessing at most~$\ell_2$ entries from~$\boldy_i$. Since $$\boldw\boldx^\intercal=\sum_{i=1}^{t(u+v)}\boldw_i\boldx_i^\intercal,$$ access requirement is at most $\ell=tu\ell_1+tv\ell_2$.
  The resulting access ratio is clearly $$\ell/k=\frac{u \ell_1+v \ell_2}{u k_1+v k_2}.$$ 
 \end{proof}
 \begin{corollary}
     Suppose there exist two $\{\pm1\}$-protocols $P1$ and $P2$ with finite parameters $(k_1,n_1,\ell_1)$ and $(k_2,n_2,\ell_2)$ respectively. Then for every rational number $\lambda$ such that $0 \le \lambda \le 1$, the pair~$\Big(\lambda\frac{n_1}{k_1}+(1-\lambda)\frac{n_2}{k_2}, \lambda\frac{\ell_1}{k_1}+(1-\lambda)\frac{\ell_2}{k_2} \Big)$ is $\{\pm1\}$-feasible.
 \end{corollary}
 \begin{proof}
 Since $\lambda$ is rational number such that $0 \le \lambda \le 1$, there exist two integers $u',v'$ such that $\lambda=\frac{u'}{u'+v'}$. Let $\zeta=\text{lcm}\{k_1,k_2\}$. Setting $u=\frac{u'\zeta}{k_1}$ and $v=\frac{v'\zeta}{k_2}$ in the feasible pair expression given in Proposition~\ref{prop:combination} results in the pair~$\Big(\frac{u'}{u'+v'}\frac{n_1}{k_1}+\frac{v'}{u'+v'}\frac{n_2}{k_2}, \frac{u'}{u'+v'}\frac{\ell_1}{k_1}+\frac{v'}{u'+v'}\frac{\ell_2}{k_2} \Big)$. 
 \end{proof}
	\section{General Quantized Linear  Computations}\label{section:general}
	Before describing our universal protocol for general quantized linear computations, we first present a new additive combinatorial notion of complexity of a finite subset of~$\bR$. The connection between this additive complexity definition and the feasible pairs of our universal protocol is established in Theorem~\ref{theorem:universal} which follows. 
	
	\begin{definition} \label{defn:complexity1}
		For any finite set $\cA \subset \bR$, the (additive) complexity of $\cA$, denoted by $C(\cA)$, is defined as the smallest positive integer~$\theta$ for which there exist $2\theta$ (not necessarily distinct) real numbers $x_1, \dots,x_{\theta}, y_1,\dots,y_{\theta}$  such that 
		\begin{equation*}
			\cA \subseteq
			\smallplus_{i=1}^{\theta} \{x_i,y_i\}. 
		\end{equation*}
	\end{definition}
	\begin{example} \label{example1}
		For example, $C(\{1,2,3,4\})=2$ as $$\{1,2,3,4\}=\{1,3\}+\{0,1\}.$$
		Now consider the set $\{1,2,3,5\}$. It can be easily proved (see Proposition~\ref{prop:set4} in the next section) that there are no four real numbers $x_1,x_2,y_1,y_2$ such that 
		$$\{1,2,3,5\} \subseteq \{x_1,y_1\}+\{x_2,y_2\}.$$
		Therefore, $C(\{1,2,3,5\})=3$ as $$\{1,2,3,5\} \subset \{0,1\}+\{0,2\}+\{0,2\}.$$	
	\end{example} 
	We now provide upper and lower bounds on the complexity of a set.  
	\begin{lemma} \label{lem:complexity_bounds}
		For any finite set $\cA$ of size at least two, 
		\begin{equation*}
			\lceil \log_2(|\cA|) \rceil  \le C(\cA) \le |\cA|-1. 	
		\end{equation*}
	\end{lemma}
	\begin{proof}
		For any collection of $2\theta$ real numbers $x_1, \dots,x_{\theta}, y_1,\dots,y_{\theta}$, we have
		\begin{equation*}
			\left|\smallplus_{i=1}^{\theta} \{x_i,y_i\}\right| \le 2^{\theta}.  
		\end{equation*}
		It follows that $$|\cA| \le 2^{C(\cA)},$$ which leads to the required lower bound.   
		
		For the upper bound, suppose $|\cA|=M$ and let the entries be given by $\cA=\{a_1,a_2,\dots,a_M\}$.  It can be verified that  
		\begin{equation*}
			\{a_1,a_2,\dots,a_M\} \subseteq  \smallplus_{i=1}^{M-2}\{a_i-a_M,0\}+\{a_{M-1},a_M\}, 
		\end{equation*}
		from which the upper bound follows. 
	\end{proof}
	\begin{example} 
		Consider the sets in Example~\ref{example1}.	The set $\{1,2,3,4\}$ is of size $4$ and complexity $2$, which is the lowest possible complexity for a set of size~$4$. On the other hand, $\{1,2,3,5\}$ is a set of size $4$ and complexity $3$, which is the highest possible complexity for a set of size~$4$.
	\end{example}  
	We will show in the next section that these bounds are tight in the sense that for every positive integer greater than one there exist sets of this size with complexity equal to the lower bound and the upper bound. We turn to show the connection between~$C(\cA)$ and low-access computation of~$\cF_\cA$, for which the following lemma is given.
	\begin{lemma}\label{lemma:complexity}
		Let $\cA \subset \bR$ be any finite set and $k \ge |\cA|$. The complexity of $\cA$ is at most $\theta$ if and only if every $\boldw \in \cA^k$  can be written as $\boldw= \sum_{i=1}^{\theta} \boldw^{(i)}$,
		where each $ \boldw^{(i)}$ is at most two-valued. 
	\end{lemma}
	\begin{proof}
		Suppose $C(\cA) \le \theta$, then it is possible to write 
		$$\cA \subseteq \smallplus_{i=1}^{\theta} \{x_i,y_i\}.$$ 
		It follows that for every $\boldw \in \cA^k$, there exists $\boldw^{(i)} \in \{x_i,y_i\}^k$ for all $i \in [\theta]$ such that 
		$\boldw= \sum_{i=1}^{\theta} \boldw^{(i)},$ thereby proving the only if part. 
		
		To prove the other direction, let $\boldw \in \cA^k$ be such that for all $a \in \cA$ there exists some $w_i=a$.  Suppose $\boldw= \sum_{i=1}^{\theta} \boldw^{(i)}$ with $\boldw^{(i)} \in \{x_i,y_i\}^k$ for some~$x_i,y_i$ and for all $i \in [\theta]$. Then, $$a \in \smallplus_{i=1}^{\theta} \{x_i,y_i\}$$ for all $a \in \cA$, from which we have $C(\cA)
		\le \theta$.
	\end{proof}

	We are ready to present our universal protocol and its connection to the complexity notion defined above. 
	\begin{theorem}\label{theorem:universal}
		If $(\alpha,\beta)$ is $\{\pm 1\}$-feasible, then $(\alpha,C(\cA)\beta)$ is $\cA$-feasible using the same encoding scheme for all finite sets $\cA \subset \bR$, i.e., a universal protocol. 
	\end{theorem}
	\begin{proof}
		Given a $\{\pm1\}$-protocol, store~$\boldx$ according to it, with an additional server containing~$\1 \boldx^\intercal$ (if not already present), where~$\1$ is the all~$1$'s vector. Let $\cA$ be any finite subset of real  numbers and let $\theta=C(\cA)$.
		Then, given any~$\boldw\in \cA^k$, by Lemma~\ref{lemma:complexity} there exist $2\theta$ real numbers $x_1, \dots,x_{\theta}, y_1,\dots,y_{\theta}$ such that $\boldw^{(i)} \in \{x_i,y_i\}^k$ and 
		\begin{equation*}
			\boldw= \sum_{i=1}^{\theta} \boldw^{(i)}. 
		\end{equation*}
		For all $i \in [\theta]$, define $\hat{\boldw}^{(i)} \in \{\pm1\}^k$ as 
		\begin{equation} \label{eq:pm1_breakdown}
			\hat{w}^{(i)}_j= \begin{cases}
				\phantom{+}1 ~\text{if}~w^{(i)}_j=x_i	\\ -1 ~\text{if}~w^{(i)}_j=y_i
			\end{cases}. 
		\end{equation}
		Then, using the $\{\pm 1\}$-protocol for retrieving $\hat{\boldw}^{(i)}$ for all $i \in [\theta]$ and directly accessing $\1\boldx^\intercal$ from its designated server, the user can compute
		\begin{align*}
			&\sum_{i=1}^{\theta}\tfrac{x_i-y_i}{2}\hat{\boldw}^{(i)}\boldx^\intercal+\Big(\sum_{i=1}^{\theta}\tfrac{x_i+y_i}{2}\Big)\1\boldx^\intercal\\
			&=\sum_{i=1}^{\theta} \Big(\tfrac{x_i-y_i}{2}\hat{\boldw}^{(i)}+\tfrac{x_i+y_i}{2}\1\Big)\boldx^\intercal\\
			&=	\sum_{i=1}^{\theta} \boldw^{(i)}\boldx^\intercal=\boldw\boldx^\intercal.
		\end{align*}
		Note that irrespective of the coefficient set $\cA$, the encoding scheme remains the same as the underlying~$\{\pm1\}$-protocol up to at most one additional server, and the access is at most $C(\cA)$ times the access of the $\{\pm 1\}$-protocol plus one. Since the feasibility of any~$(\alpha,\beta)$ is defined asymptotically, the claim follows.
	\end{proof}
	From Proposition~\ref{theorem:universal} and the complexity upper bound in Lemma~\ref{lemma:complexity}, we can always guarantee the following. 
	\begin{corollary}\label{corollary:maxAccess}
		If $(\alpha,\beta)$ is $\{\pm 1\}$-feasible, then $(\alpha,(|\cA|-1)\beta)$ is $\cA$-feasible using the same encoding scheme for all finite sets $\cA \subset \bR$. 
	\end{corollary}
	
	From Theorem~\ref{theorem:universal} and Corollary~\ref{corollary:maxAccess}, it follows that the~$\{\pm 1\}$-protocol (which is also an~$\cF_2$-protocol) can be used as a black-box for the retrieval of~$\cF_\cA$ for any finite~$\cA$, with no penalty on the asymptotic storage ratio, and the aforementioned multiplicative penalty on the access ratio. Furthermore, it is possible to get an access ratio better than the one given in Corollary~\ref{corollary:maxAccess} if the coefficient set is not of the highest  complexity. Getting a good handle on complexity will help us to understand the implications of Theorem~\ref{theorem:universal} better, as the access ratio of our universal protocol depends on the complexity of the coefficient set. The rest of the paper studies various properties of the complexity of sets. 

	\section{Additive Complexity of a set}\label{section:complexity}
	In this section, we explore the new notion of additive complexity of a set in more detail. We  first characterize the complexity of sets of size $3$ and $4$. For sets of larger size, characterizing the complexity seems to be a difficult task, and yet we are able to provide infinite examples of sets of highest and lowest complexity. A connection between our
	complexity definition and the well-known number-theoretic notion of Sidon sets is also established.
	
	\subsection{Small sets and basic properties}
	\begin{proposition}
		All sets of size $3$ have complexity $2$. 
	\end{proposition}
	\begin{proof}  Follows  from Lemma~\ref{lem:complexity_bounds}. 
	\end{proof} 
	\begin{corollary}
		If $(\alpha,\beta)$ is $\{\pm 1\}$-feasible, then $(\alpha,2\beta)$ is $\{a,b,c\}$-feasible using the same encoding scheme, for all distinct $a,b,c \in \bR$. 
	\end{corollary}
	\begin{proposition} \label{prop:set4}
		Sets of size $4$ have complexity $2$ if the sum of the largest element and the smallest element is equal to the sum of the other two elements. Sets of size $4$ not satisfying this condition have complexity $3$. 
	\end{proposition}
	\begin{proof}
		Let $\cA=\{a,b,c,d\}$  with $a<b<c<d$. By Lemma~\ref{lem:complexity_bounds}, we have $2 \le C(\cA) \le 3$. If $a+d=b+c$, then we can write
		$$\cA=\{a,b\} +\{c-a,0\},$$ and hence $C(\cA)=2$ for this case. 
	  
  Now assume that $C(\cA)=2$ and   $\cA  \subseteq \{x_1,y_1\} + \{x_2,y_2\}$, where $x_i<y_i$. Since $\cA$ is of size~$4$, it follows that $$\cA = \{x_1,y_1\} + \{x_2,y_2\}=\{x_1+x_2,x_1+y_2,y_1+x_2,y_1+y_2\}.$$
  Clearly, the largest and smallest elements are  $y_1+y_2$ and $x_1+x_2$, respectively, and it is easy to verify that their sum is equal to the sum of the other two elements. 
  Hence, if the set $\cA$ satisfies $a+d\neq b+c$, then necessarily, $C(\cA)=3$.
  

	\end{proof}
	\begin{corollary}
		If $(\alpha,\beta)$ is $\{\pm 1\}$-feasible, then $(\alpha,2\beta)$ is $\{a,b,c,d\}$-feasible using the same encoding scheme for all real numbers $a<b<c<d$ with $a+d=b+c$, and~$(\alpha,3\beta)$ is~$\{a,b,c,d\}$-feasible using the same encoding scheme for all real numbers~$a<b<c<d$ with~$a+d\ne b+c$.
	\end{corollary}
	We now present some basic properties of complexity. 
	\begin{lemma}
		The complexity of a finite set does not change if it is shifted or scaled. More formally, for every finite set $\cA$, 
		\begin{itemize}
			\item $C(\cA)=C(\cA+\{s\})$ for all $s \in \bR$,
			\item $C(\cA)=C(s\cA)$ for all $s \in \bR \setminus \{0\}$. 
		\end{itemize}
	\end{lemma}
	\begin{proof}
		Suppose $\cA \subseteq \smallplus_{i=1}^{\theta} \{x_i,y_i\}$. Then,  $$\cA + \{s\} \subseteq \smallplus_{i=1}^{\theta} \big\{x_i+\frac{s}{\theta},y_i+\frac{s}{\theta}\big\}$$
		and 
		$$s\cA\subseteq \smallplus_{i=1}^{\theta} \big\{sx_i,sy_i\big\}.$$
		It is easy to see that the statement of lemma follows. 
	\end{proof} 
	\subsection{Equivalent definitions}
	
	We now give two alternative definitions of complexity that will be shown to be equivalent. The first one gives more information on the structure of the component sets, whereas the second definition is an algebraic viewpoint. 

	\begin{definition} \label{defn:complexity2}
		For any finite set $\cA \subset \bR$, the complexity of $\cA$ is the smallest positive integer $\theta$ for which there exist $\theta$ positive real numbers $z_1, \dots,z_{\theta}$ and a (not necessarily positive) real number $s$ such that 
		\begin{equation*}
			\cA \subseteq \smallplus_{i=1}^{\theta} \{z_i,0\}+\{s\}. 
		\end{equation*}
	\end{definition}
	\begin{example}
		We revisit the sets in Example~\ref{example1}. Recall the $C(\{1,2,3,4\}) = 2$ and $C(\{1,2,3,5\}) = 3$. It can be verified that
		$$\{1,2,3,4\}=\{0,2\}+\{0,1\}+\{1\},$$ and
		$$\{1,2,3,5\} \subset \{0,1\}+\{0,2\}+\{0,2\}+\{0\}.$$
	\end{example} 
	For any finite set $\cA \subset \bR$,  we will use 
	$\operatorname{vec}(\cA)$ to denote the $|\cA| \times 1$ column vector obtained by listing the entries of $\cA$ in increasing order. The exact ordering is irrelevant to our discussion, and we assume without loss of generality that it is an increasing order.   
	\begin{definition} \label{defn:complexity3}
		The complexity of a finite set $\cA \subset \bR$ is the smallest positive integer $\theta$ for which there exists a  $\{0,1\}$ matrix $B$ of size $|\cA| \times (\theta+1)$ such that
		\begin{itemize}
			\item $B$ has an all~$1$'s column; and
			\item  $\operatorname{vec}(\cA)$ lies in the column span (over $\bR$) of $B$.
		\end{itemize}
	\end{definition}
	\begin{example}
		For $\cA=\{1,2,3,4\}$, we have $$\operatorname{vec}(\cA)=[1~2~3 ~4]^\intercal$$ and   
		$$ \underbrace{ \begin{bmatrix}
				1 & 0 & 0 \\
				1 & 0 & 1 \\
				1 & 1 & 0 \\
				1 & 1 & 1 \\
		\end{bmatrix} }_{B} \begin{bmatrix}
			1 \\
			2 \\
			1 \\
		\end{bmatrix}=  \begin{bmatrix}
			1 \\
			2 \\
			3 \\
			4
		\end{bmatrix}.$$  
		For $\cA=\{1,2,3,5\}$, we have $$\operatorname{vec}(\cA)=[1~2~3 ~5]^\intercal$$ and 
		$$  \underbrace{  \begin{bmatrix}
				1 & 1 & 0 & 0\\
				1 & 0 & 1 & 0\\
				1 & 1 & 1 & 0\\
				1 & 1 & 1 & 1\\
		\end{bmatrix}}_{B}  \begin{bmatrix}
			0 \\
			1 \\
			2 \\
			2
		\end{bmatrix}=  \begin{bmatrix}
			1 \\
			2 \\
			3 \\
			5
		\end{bmatrix}.$$  
	\end{example}
	\begin{proposition}\label{proposition:equivalence}
		All three definitions of complexity are equivalent. 
	\end{proposition}
	\begin{proof} First we prove that Definition~\ref{defn:complexity1} and Definition~\ref{defn:complexity2} are equivalent. Suppose $\cA \subseteq \smallplus_{i=1}^{\theta} \{x_i,y_i\}$, with $x_i \ne y_i$ for $i\in[\theta]$. 
		Each $a\in\cA$ can be written as a summation of~$\theta$ elements, one from each set~$\{x_i,y_i\}$ for~$i\in[\theta]$. Let~$I_a\subseteq[\theta]$ be the set of indices such that the larger among~$\{x_i,y_i\}$ shows up in the above representation of~$a$, and let~$J_a$ be its complement. That is, we have
		$$a=\sum_{i \in I_a}\max\{x_i,y_i\} + \sum_{j \in J_a}\min\{x_i,y_i\}.$$	Set $$z_i = \max\{x_i,y_i\}-\min\{x_i,y_i\},$$
		for all $i \in [\theta]$ and $$s= \sum_{i=1}^{\theta}\min\{x_i,y_i\}.$$
		It follows that  
		$$\sum_{i \in I_a}z_i + s= \sum_{i \in I_a}\max\{x_i,y_i\} + \sum_{j \in J_a}\min\{x_i,y_i\}=a.$$
		Hence, $ \cA \subseteq \smallplus_{i=1}^{\theta} \{z_i,0\}+\{s\}$ and all $z_i$ are positive. The other direction also holds since $$\smallplus_{i=1}^{\theta} \{z_i,0\}+\{s\}=\smallplus_{i=1}^{\theta} \Big\{z_i+\frac{s}{\theta},\frac{s}{\theta}\Big\}.$$
		
		Now we show that Definition~\ref{defn:complexity2} and Definition~\ref{defn:complexity3} are equivalent.  This is true since $$\cA \subseteq \smallplus_{i=1}^{\theta} \{z_i,0\}+\{s\}$$ if and only if there exist binary column vectors $\underbar{b}_1, \cdots,\underbar{b}_{\theta}$ such that
		\begin{align}
			\operatorname{vec}(\cA)&=s \1^\intercal+ \sum_{i=1}^{\theta}z_i\underbar{b}_i. \qedhere
		\end{align}
	\end{proof}
	\subsection{Sets of lowest complexity}
	Next, examples of infinite families of sets with the lowest possible complexity are presented.
	\begin{proposition}
		For every integer $M \ge 2$, there exist sets of size $M$ with complexity $\lceil \log_2(M)\rceil$, which is the lowest possible. Furthermore, arithmetic progressions are examples of such sets. 
	\end{proposition}
	\begin{proof} 
		Consider a set $\cA \subset \bR$ of size $M$ such that the elements are in arithmetic progression, i.e., $$\cA=\{a+ix \mid 0 \le i  \le M-1  \},$$
		where $x > 0$. 
		By looking at the base $2$ representation of integers it can be verified that
		$$ \smallplus_{j=0}^{\lceil \log_2(M)\rceil-1} \{2^jx,0\} + \{a\}=\{a+ix \mid 0 \le i  \le M^*  \},$$
		where 
		$M^*=\sum_{j=0}^{\lceil \log_2(M)\rceil-1}2^j=2^{\lceil \log_2(M)\rceil}-1$. Clearly, $M-1 \le M^*$, and hence 
		$$ \cA \subseteq \smallplus_{j=0}^{\lceil \log_2(M)\rceil-1} \{2^jx,0\} + \{a\}.$$ By Definition~\ref{defn:complexity2} and the equivalence in Proposition~\ref{proposition:equivalence}, it follows that
		$C(\cA) \le \lceil \log_2(M)\rceil$ and the lower bound in Lemma~\ref{lem:complexity_bounds} makes it an equality.  	
	\end{proof}
		
	\begin{remark}
		Uniform quantization~\cite[Chapter~3]{gallager2008principles} is often preferred in practical settings due to its ease of implementation, and results in equally spaced quantization levels, i.e., an arithmetic progression. It follows from the above corollary that our universal protocol works well with uniform quantization, due to the resulting lowest possible complexity.
	\end{remark}
	\subsection{Sets of highest complexity}
	
	Next, sets with the highest possible complexity are considered. The performance of our universal protocol is not good for such sets.  
	We show below in Proposition~\ref{prop:allbutzero} that most sets have the highest possible complexity. If access is a major concern, one needs to be careful to avoid such sets while performing the quantization.  Before stating Proposition~\ref{prop:allbutzero}, we first introduce some terminology. Let $\bR^M_{*}$ be the subset of $\bR^M$ containing vectors whose entries are in a strictly increasing order. It can be seen that $\bR^M_{*}$ is the intersection of a finite number of open half-spaces in $\bR^M$, i.e., an open convex polytope, and hence $\bR^M_{*}$ has non-zero measure in $\bR^M$. For  every vector $\boldv \in \bR^M_{*}$, there is an underlying set $\cA$ of size $M$ such that $\operatorname{vec}(\cA)=\boldv^\intercal$. 
	
	\begin{proposition} \label{prop:allbutzero}
 All vectors in $\bR^M_{*}$, but a set of measure zero, have underlying sets of complexity $M-1$, i.e., the highest possible complexity.
	\end{proposition}
	\begin{proof} 
		Consider any set $\cA \subset \bR$ of size $M$ with $C(\cA)<M-1$. Then, by Definition~\ref{defn:complexity3} and the equivalence in Proposition~\ref{proposition:equivalence}, there exists an $M \times (M-1)$ binary matrix $B$ such that $\operatorname{vec}(\cA)$ is in the column span of $B$. In this proof, we consider vectors in $\bR^M$ as column vectors. The column span of every $M \times (M-1)$ matrix is contained in a hyperplane in $\bR^M$. There are finitely many $M \times (M-1)$ binary matrices. The union of finitely many hyperplanes has measure zero in $\bR^M$
  as well as in $\bR^M_{*}$.
	\end{proof}
	We now show an example family of sets with the highest possible complexity.  
	\begin{proposition}
		Geometric progressions with sufficiently large common ratios have the highest possible complexity.   
	\end{proposition}
	\begin{proof}
		Consider a set $\cA \subset \bR$ of size $M$ such that the elements are in geometric progression with a common ratio greater than one, i.e., $$\cA=\{ar^i \mid 0 \le i  \le M-1 \},$$
		where $a \ne 0$ and $r>1$. 
		Our aim is to prove that $C(\cA)=M-1$ by induction, for large $r$. Clearly, the base case holds as $C(\{a,ar\})=1$.  The induction assumption is  $C(\cA')=M-2$, where
		$\cA'=\{ar^i \mid 0 \le i  \le M-2 \}$. 
		
		
		Suppose $C(\cA) \le M-2$.  
		Then, by Definition~\ref{defn:complexity3} and the equivalence in Proposition~\ref{proposition:equivalence}, there exists a matrix $B \in \{0,1\}^{M \times (M-1)}$ such that $\operatorname{vec}(\cA)$ is in the column span of $B$. 	Additionally, we can assume that the first column of $B$  is an all $1$'s column.  Let $\boldb_1, \boldb_2, \dots,\boldb_M$ denote the rows of $B$. 
		$$ B = \begin{bmatrix}
			\boldb_1 \\
			\boldb_2 \\
			\vdots \\
			\boldb_M
		\end{bmatrix}.$$
		Let $\hat{B}$ be the $M \times M$ matrix obtained by augmenting  the column $\operatorname{vec}(\cA)$ to $B$, i.e., 
		$$ \hat{B} = \begin{bmatrix}
			B & \operatorname{vec}(\cA)
		\end{bmatrix}.$$
		Since $\operatorname{vec}(\cA)$ lies in the column span of $B$, we have
		\begin{equation}
			\label{eq:det_zero}
			\det(\hat{B})=0.
		\end{equation}
		For every $i \in [M]$, let $B_i$ denote the $(M-1) \times (M-1)$ matrix obtained by deleting row $\boldb_i$ from $B$, i.e.,
		$$ B_i = \begin{bmatrix}
			\boldb_1 \\
			\vdots  \\
			\boldb_{i-1} \\
			\boldb_{i+1} \\
			\vdots  \\
			\boldb_M
		\end{bmatrix}$$
		It follows from the well-known Hadamard's inequality that the determinant of any $\{0,1\}$ square matrix of order $t$ is upper bounded by $$D(t) \triangleq \frac{t^{\frac{t}{2}}}{2^{t-1}}.$$ Hence for all $i \in [M]$, it holds that
		\begin{equation}
			\label{eq:det_upperbound}
			\det(B_i) \le D(M-1). 
		\end{equation} 
		We claim that
		\begin{equation} \label{eq:Bn_det}
			|\det(B_M)| \ge 1; 
		\end{equation}
		the proof of this claim appears after the proof of the current proposition. From \eqref{eq:det_zero}, we have $ |\det(\hat{B})|=0$. Expanding the determinant of $\hat{B}$ along the last column leads to,
		$$\Bigg{|}\sum_{i=0}^{M-1}(-1)^{i-M+1} ar^{i} \det(B_{i+1})\Bigg{|}=0.$$
		Since $a \ne 0$ and $|x+y| \ge |x|-|y|$, the above equation implies
		$$\big{|} r^{M-1} \det(B_M)\big{|}-\sum_{i=1}^{M-1} \big{|} r^{i-1} \det(B_i)\big{|} \le 0.$$ 
		
		It follows from ~\eqref{eq:det_upperbound},~\eqref{eq:Bn_det}, and $r>0$ that the above inequality results in 
		\begin{align}
			r^{M-1}-D(M-1)\sum_{i=1}^{M-1} r^{i-1} &\le 0 \nonumber \mbox{, and hence}\\
			  r^{M-1}-D(M-1) \frac{r^{M-1}-1}{r-1} &\le 0. \label{eq:contradiction}
		\end{align}
		If $r$ is sufficiently large, for instance, if
		$$ r \ge D(M-1)+1,$$
		it follows that the inequality \eqref{eq:contradiction} is false,  resulting in a contradiction.  Hence, $C(\cA) > M-2$, and the statement of the proposition follows.  
	\end{proof}
	\begin{proof}[Proof of the claim $|\det(B_M)| \ge 1$]
		Since the entries of $B_M$ are in $\{0,1\}$, to prove the claim it suffices to show that $\det(B_M) \ne 0$. Denote
		$$B_M	=  \begin{bmatrix}
			\underline{c}_1 &
			\underline{c}_2 &
			\dots &
			\underline{c}_{M-1}
		\end{bmatrix}.$$
		If $\det(B_M) =0$, then there exists a $u \in [M-1] \setminus \{1\}$ and $\{\lambda_j\}_{j \in [M-1], j \ne u} \subset \bR $ such that
		\begin{align}
			\label{eq:lin_dep}
			\underline{c}_{u} = \sum_{j \in [M-1], j \ne u} \lambda_j \underline{c}_j.
		\end{align}
		Let $B'_M$ be the $(M-1) \times (M-2)$ matrix obtained by deleting column $\underline{c}_{u}$ from $B_M$, i.e., 
		$$B'_M	=  \begin{bmatrix}
			\underline{c}_1 &
			\cdots &
			\underline{c}_{u-1} &
			\underline{c}_{u+1} &
			\cdots &
			\underline{c}_{M-1}
		\end{bmatrix}.$$
		Since $\operatorname{vec}(\cA)$ is in the column span of $B$, it follows that $\operatorname{vec}(\cA')$ is in column span of $B_M$.  Due to  \eqref{eq:lin_dep}, the column spans of $B_M$ and $B'_M$ are identical, and therefore $\operatorname{vec}(\cA')$ is also in the column span of $B'_M$. Since $u \ne 1$, $B'_M$ contains an all~$1$'s  column. Then by Definition \ref{defn:complexity3}, it must hold that $C(\cA') \le M-3$. However, this contradicts our induction assumption that $C(\cA')=M-2$. Hence, $\det(B_M)  \ne 0$. 
	\end{proof}
	

We now turn to show that even without a large common ratio, geometric progressions are most likely not a good choice for low-access linear computation. In Proposition~\ref{prop:allbutzero}, we showed that all but a measure-zero set of vectors correspond to the highest complexity sets; next we show that a similar phenomenon holds if one is restricted to geometric progressions.
	
	\begin{proposition} 
 \label{prop-9}
For any $M\geq 3$ and any $a \ne0$, all but a finite number of the geometric progressions $\{ar^i \mid 0 \le i  \le M-1 \}$ of length $M$ have the highest possible complexity.
	\end{proposition} 	
	\begin{proof}
		Consider a set $\cA \subset \bR$ of size $M \ge 3$ such that the elements are in geometric progression, i.e., $$\cA=\{ar^i \mid 0 \le i  \le M-1 \},$$
		where $a \ne 0$ and $r \notin \{0,\pm 1\}$. Note that $r\notin \{0,\pm 1\}$ because otherwise, the set $\cA$ would be of size at most $2$. Suppose $C(\cA)<M-1$. Then by Definition~\ref{defn:complexity3}, there exists a matrix $B \in \{0,1\}^{M \times (M-1)}$ such that $\operatorname{vec}(\cA)$ is in the column span of $B$. 
		
		Let $\operatorname{rank}(B) = \lambda \le M-1$. It follows that $B$ has an $\lambda \times \lambda$ invertible submatrix, we call it $B_{inv}$.  Let $S \subseteq [M]$ and $T \subseteq [M-1]$, with $|S|=|T|=\lambda$, be such that $B_{inv}$ is formed by the entries of~$B$ with row index in~$S$ and column index in~$T$. Pick an $S' \subseteq [M]$ such that $S \subset S'$ and $|S'|=\lambda+1$. Note that $S' \setminus S$ contains a single element and let it be $\beta$.
		
		Let $B'$ be the $(\lambda+1) \times \lambda$ sub-matrix of $B$ formed by entries of $B$ with row index in $S'$  and column index in $T$, and let $\cA' = \{ar^{i-1} \mid i \in S'\}$. Let $\hat{B}$ be the $(\lambda+1) \times (\lambda+1)$ matrix obtained by augmenting the column $\operatorname{vec}(\cA')$ to $B'$, i.e., 
		$$ \hat{B} = \begin{bmatrix}
			B' & \operatorname{vec}(\cA')
		\end{bmatrix}.$$
		By the property of $B$, it  must hold that $\operatorname{vec}(\cA')$ is in the column span of $B'$, and hence $\det(\hat{B})=0$. For any fixed $a \ne 0$, $\det(\hat{B})$ can be written as a polynomial in~$r$ of degree at most $M-1$, by expanding along the last column of $\hat{B}$. 
 Since $\det(B_{inv}) \ne 0$, 
  the coefficient of $r^{\beta}$ in this polynomial is non-zero and hence it is a non-zero polynomial. Therefore, for any given real number $a \ne 0$, there are at most $M-1$ possible real values for~$r$ which result in $\det(\hat{B})=0$. The number of distinct $M \times (M-1)$ binary matrices is $2^{M(M-1)}$, i.e., finite. Hence, for all non-zero real numbers~$a$, except for a finitely many  $r$, the complexity of $\{ar^i \mid 0 \le i  \le M-1 \}$ is $M-1$.
	\end{proof}
Note that the proof of Proposition~\ref{prop-9} does not imply the existence of a single geometric progression without the highest possible complexity. The next proposition, however, shows that not all geometric progressions have the highest possible complexity. 
	\begin{proposition}
		For every integer $M \ge 5$, there exist $M$ element geometric progressions which do not have the highest possible complexity.   
	\end{proposition}
	\begin{proof} We begin by presenting a sufficient condition for the existence of a geometric progression set of size $M$ and complexity less than $M-1$.  Suppose there exists an $M \times (M-1)$ binary matrix $B$ having the following properties: 
		\begin{itemize}
			\item $\operatorname{rank}(B)=M-1$; 
			\item 	 $B$ contains the all~$1$'s column; and 
			\item there exists a real number $x \notin \{0,1,-1\}$ for which the $M \times M$ matrix 
			$$ B_x = \begin{bmatrix}
				B & \boldx^\intercal
			\end{bmatrix}$$
			is not invertible, where $\boldx=[1~x~x^2~\dots~x^{M-1}]$. 
		\end{itemize}
		Then by Definition \ref{defn:complexity3}, the complexity of $\{1,x,x^2,\dots,x^{M-1}\}$ is at most $M-2$.
		Note that $\{1,x,x^2,\dots,x^{M-1}\}$  is a set of $M$ elements in geometric progression. Thus, to prove the result it is sufficient to show the existence of such  $B$ matrices for all $M \ge 5$.  
		
		We prove this by induction on the size of the geometric progression set. As base case, for size $5$, the $5 \times 4$ matrix 
		$$ \begin{bmatrix}
			1 & 1 & 1 & 0\\
			1 & 1 & 0 & 1\\
			1 & 1 & 0 & 0\\
			1 & 0 & 1 & 1\\
			1 & 0 & 0 & 0
		\end{bmatrix}$$
		satisfies the required properties, as $\det(B_x)=0$ if $x=\frac{1+\sqrt{5}}{2}$, which is the golden ratio.
		
		The induction assumption is the existence of a $(\mu-1) \times (\mu-2)$ binary matrix $B$ satisfying the above properties for size $\mu-1>5$. Without loss of generality, let the first column of $B$ be the all~$1$'s column. It can be verified  that the $\mu \times (\mu-1)$ binary matrix 
		$$
		\left[ \begin{array}{ccccc}
			\multicolumn{4}{c}{ \multirow{4}{*}{$B$}} & 0 \\
			& & & & \vdots \\
			& & & & 0 \\
			1	& 0 & \dots & 0 & 1 \\
		\end{array}\right],
		\label{eq:H_example}
		$$
		satisfies all the required properties for size $\mu$, thereby completing the proof.
	\end{proof}
	\subsection{Connection to Sidon sets}
	
 Here, we show a connection between our definition of complexity and a well-known notion in number theory.
 \begin{definition}[\cite{o2004complete}]
		A finite set $\cA=\{a_1,\dots,a_M\} \subset \bR$ of size $M$ is a Sidon set if all pairwise sums $a_i+a_j$, with $i,j \in [M]$ and $i \le j$, are different\footnote{Strictly speaking, Sidon sets are contained in~$\bN$, however in this paper we consider them as contained in~$\bR$.}. 
	\end{definition} 
We now introduce a slightly relaxed definition, which we call almost-Sidon sets, where the sum of an element with itself is allowed to match another pairwise sum.  
\begin{definition}
		A finite set $\cA=\{a_1,\dots,a_M\} \subset \bR$ of size $M$ is an almost-Sidon set if all pairwise sums $a_i+a_j$, with $i,j \in [M]$ and $i < j$, are different. 
	\end{definition} 
 For example, $\{0,1,2\}$ is not a Sidon set as $0+2=1+1$, but it is an almost-Sidon set. On the other hand, $\{0,1,3\}$ is  a Sidon set and therefore also an almost-Sidon set. 
 The next proposition provides a necessary condition for a set to be of the highest complexity.  
	\begin{proposition} Every set of the highest possible complexity is an almost-Sidon set.
 \end{proposition}
	\begin{proof}
		Suppose $\cA=\{a_1,\dots,a_M\}$ is not almost-Sidon set, i.e., there exist distinct indices $i_1,i_2,j_1,j_2 \in [M]$ such that  $$a_{i_1}+a_{j_1}=a_{i_2}+a_{j_2}.$$ Without loss of generality one can  assume $a_{j_1}>a_{i_2}$. Define 
		$\cA'=\cA \setminus \{a_{j_1},a_{j_2}\}$.   Then, we have
		$$ \cA \subseteq \cA' + \{a_{j_1}-a_{i_2},0\},$$
		as  $$a_{j_1}=a_{i_2}+(a_{j_1}-a_{i_2})~~\text{and}~~ a_{j_2}=a_{i_1}+(a_{j_1}-a_{i_2}).$$ 
		It follows from Definition~\ref{defn:complexity2} that 
		$$C(\cA) \le C(\cA')+1.$$ 
		Since $\cA'$ is of size $M-2$, by the upper bound in Lemma~\ref{lem:complexity_bounds}, we get 
		$C(\cA') \le M-3.$
		Hence,
		$C(\cA) \le M-2,$
		thereby proving the proposition. 
  \end{proof}
The above proposition gave a necessary condition for a set to be of the highest complexity; then it would be natural to ask whether this is also a sufficient condition.  The following proposition refutes this in the strongest sense, i.e., there exist almost-Sidon sets whose complexity up to a constant factor is the smallest possible.
   \begin{proposition}
There exist almost-Sidon sets of size $M$ with complexity $O(\log_2M)$ for infinitely many values of~$M$. 
	\end{proposition} 
  \begin{proof}
	Let $H$ be a $(2m+1)\times (2^m-1)$ parity check matrix of a binary BCH codes having parameters $[2^m-1, 2^m-2m-1, d_{\min} \ge 5]$
  for some $m\geq3$ (see \cite[Chapter~7]{ecc_Mac_Slo}).
 Set $M= 2^m-1$ and let $B$ be the $M \times (2m+2)$ binary matrix (over $\mathbb{R}$) obtained by appending an all~$1$'s column to the transpose of $H$, i.e.,
		$$ B = \begin{bmatrix}
			H^\intercal & \1^\intercal
		\end{bmatrix}.$$
		Let $L=2m+2$ denote the number of columns of $B$, and let  $\boldb_1, \boldb_2, \dots,\boldb_M$ be the rows of $B$, i.e., 
		$$ B = \begin{bmatrix}
			\boldb_1 \\
			\boldb_2 \\
			\vdots \\
			\boldb_M
		\end{bmatrix}.$$
Next, for any four distinct indices  $i,j,p,q \in [M]$ define the linear equations $$((\boldb_i+\boldb_j)-(\boldb_p+\boldb_q))\boldx^\intercal=0 \text{ and } (\boldb_i-\boldb_j)\boldx^\intercal=0,$$
in the variables $x_1,x_2,\dots,x_L$. 
Note that $\boldb_i-\boldb_j\neq 0$  and also $(\boldb_i+\boldb_j)-(\boldb_p-\boldb_q)\neq 0$; both of them follow by the minimum distance of the code as any $4$ columns of $H$ are linearly independent over $\bF_2$. Hence,  the solution of each equation is a hyperplane in $\bR^L$. Therefore there exists a vector $\boldx=[x_1,x_2,\dots,x_L] \in \bR^L$ that 
simultaneously does not satisfy all of the equations. Define the set $\cA$ such that  $B\boldx^\intercal=\operatorname{vec}(\cA)$. We claim that the set~$\cA$ is the required set. Indeed,  for distinct indices $i,j$ by construction 
$(\boldb_i-\boldb_j)\boldx^\intercal\neq 0$ and therefore $|\cA|=M$. Furthermore, by Definition~\ref{defn:complexity3}, the complexity of set $\cA$ is $2m+1=O(\log_2M)$. By the other set of equations, we conclude that the set $\cA$ is an almost-Sidon set. Indeed, for any four distinct elements $a,b,c,d\in \cA$
$$(a+b)-(c+d)=((\boldb_i+\boldb_j)-(\boldb_p+\boldb_q))\boldx^\intercal\neq 0,$$
where the equality follows for some distinct indices $i,j,p,q\in [M]$ and the not equal to sign follows by the choice of $x$, and the result follows.	 
	\end{proof}
 \section{Joint Computation and Generalized Covering Radius} \label{section:generalized_covering} 
 In the proof of Theorem~\ref{theorem:universal}, we showed that any function in $\cF_{\cA}$ can be broken down into $C(\cA)$ two-valued linear computations (to be precise, $\1\boldx^\intercal$ may also be needed). We then argued that it suffices to access at most $C(\cA)\ell+1$ symbols to compute any function in  $\cF_{\cA}$, where $\ell$ is the access parameter of the underlying $\{\pm 1\}$-protocol. This is achieved by separately retrieving $C(\cA)$-many two-valued linear computations. In this Section, we explore the possibility of reducing the access further, by performing $C(\cA)$ two-valued computations \textit{jointly}.
 The joint computation problem is interesting even beyond its connection to coefficient complexity,  for example, when performing inference of binarized neural networks with a large first layer, multiple $\{\pm1\}$ linear combinations need to be computed simultaneously.
 
As argued earlier in Section~\ref{section:pm1}, the pair $(1,0.5)$ is $\cF_2$-feasible. For any given finite set $\cA$, it follows by Theorem~\ref{theorem:universal} that $(1,C(\cA)/2)$ is $\cA$-feasible. However, this point of the tradeoff is useless if $C(\cA)>1$, as one can always trivially get an access ratio of one without any coding. Proposition~\ref{prop:joint} below illustrates the benefit of considering joint computation in this case. The following lemma is needed in the proof of the proposition. The lemma states that any~$\theta$ vectors in $\{\pm1\}^k$ or their negations agree on at least~$k/2^\theta$ of the entries.

\begin{lemma} \label{lemma:joint}
    Given any collection of $\theta$ vectors $\boldw^{(1)},\dots,\boldw^{({\theta})} \in \{\pm1\}^k$, there exists a set $B \subseteq [k]$ with $|B| \ge \frac{k}{2^{\theta}}$ and a vector $\bolds \in \{\pm1\}^{\theta}$ such that $s_iw_{j}^{(i)}=1$ for all $i \in [\theta]$ and $j \in B$. 
\end{lemma}
\begin{proof} We prove the lemma by induction. For the base case $\theta=1$, it is easy to see that if less than $k/2$ entries of $\boldw^{(1)}$ are $1$, then more than $k/2$ entries of $-\boldw^{(1)}$ are $1$. We now assume that the statement of the lemma is true for any collection of $\theta-1$ vectors. Given $\theta$ vectors $\boldw^{(1)},\dots,\boldw^{(\theta)} \in \{\pm1\}^k$, by the induction assumption there is a set $\hat{B} \subset[k]$ of size at least  $\frac{k}{2^{\theta-1}}$ and a vector $\hat{\bolds} \in \{\pm1\}^{\theta-1}$ such that $s_iw_{j}^{(i)}=1$ for all $i \in [\theta-1]$ and $j \in \hat{B}$. Consider the set $I=\{i \in [k]~\vert~\boldw^{(\theta)}_i=1\}$. If $|\hat{B} \cap I| \ge \frac{k}{2^{\theta}}$, then $B=\hat{B} \cap I$ and $\bolds=(\hat{\bolds},1) \in \{\pm1\}^{\theta}$ meets the requirements of the lemma. If not, $B=\hat{B} \cap ([k] \setminus I)$ and $\bolds=(\hat{\bolds},-1)$ works.   
\end{proof}
 \begin{proposition}\label{prop:joint}
The pair~$(1,1-\frac{1}{2^{C(\cA)}})$ is $\cA$-feasible, for any finite set $\cA$.  
 \end{proposition}
\begin{proof}
Store each~$x_i$ in a separate node, and include an additional node containing~$\1\boldx^\intercal$. Hence, $n=k+1$, and the redundancy ratio is one. Let $C(\cA)=\theta$ and $\boldw$ be some vector in $\cA^k$. 
Recall the definition $\hat{\boldw}^{(1)}, \dots, \hat{\boldw}^{(\theta)} \in \{\pm1\}^k$ given in~\eqref{eq:pm1_breakdown}. If the user knows  $\hat{\boldw}^{(1)}\boldx^\intercal$, \dots, $\hat{\boldw}^{(\theta)}\boldx^\intercal$ and $\1\boldx^\intercal$, it can compute $\boldw\boldx^\intercal$. It follows from   Lemma~\ref{lemma:joint} that there exists $B \subset [k]$ with $|B| \ge \frac{k}{2^{\theta}}$ and $\bolds \in  \{\pm1\}^{\theta}$ such that $s_1\hat{\boldw}^{(1)}, \dots, s_{\theta}\hat{\boldw}^{(\theta)}$ can be computed by accessing $\{x_i \vert i \in [k]\setminus B\}$ and $\1x^T$. Thus, it suffices to access $k-\frac{k}{2^{\theta}}+1$ nodes to compute $\boldw\boldx^\intercal$, resulting in access ratio of $1-\frac{1}{2^{\theta}}$.  
\end{proof} 
 
 The generalized covering radius was recently introduced in \cite{elimelech2021generalized} for applications in the joint recovery of linear computations over finite fields.  Here we explore the possibility of using codes with a small generalized covering radius for low-access joint retrieval of two-valued linear computations over reals. Proposition~\ref{prop:joint} above is equivalent to choosing the code $\cC$ in Theorem~\ref{theorem:generalzied_covering} below as the repetition code.   
 \begin{definition}[\cite{elimelech2021generalized}]
\label{defn:generalized_covering}
     Let $\cC$ be a code of length $p$ over $\bF_2$. For $\theta \in \bN$, the $\theta$-th generalized covering radius $R_\theta$ is the smallest integer $s$ such that for every $\boldv_1,\boldv_2,\dots,\boldv_\theta \in \bF_2^p$, there exist codewords $\boldc_1,\boldc_2,\dots,\boldc_{\theta} \in \cC$ such that $ |\cup_{i=1}^{\theta} \operatorname{supp}(\boldv_i-\boldc_i)| \le s$.      
 \end{definition}
 Setting $\theta=1$ in this definition gives the covering radius defined in Section~\ref{section:coverigncodes}. Furthermore, if the covering radius of~$\cC$ is~$r$, then $R_{\theta} \le \theta r$. 

 \begin{theorem}\label{theorem:generalzied_covering}
		If there exists a (not-necessarily linear) code~$\cC$ of length~$p$ over~$\bF_2$, then the pair~$(\frac{p+\hat{c}}{p},\frac{R_{C(\cA)}+C(\cA)}{p})$ is $\cA$-feasible,  using the same encoding scheme for all finite sets $\cA \subset \bR$, where $\hat{c}$ is as given by Definition \ref{definition:complementFree}. Moreover, the encoding scheme is systematic.
	\end{theorem}
 	\begin{proof}
	Construct the matrix~$M=(B\vert I)\in \{0,\pm 1\}^{p\times (p+\hat{c})}$, where~$I$ is the identity matrix and~$B$ contains all~$\hat{c}$ vectors of~$\hat{\cC}$ in their~$\{\pm 1\}$-representation as columns. For~$t\in\bN$ consider~$k=tp$, and partition~$\boldx\in\bR^{tp}$ to~$t$ parts~$\boldx_1,\ldots,\boldx_{t}$ of size~$p$ each. To encode, let~$\boldy_i\triangleq \boldx_i M$, and let~$\boldy=(\boldy_1,\ldots\boldy_t,\1\boldx^\intercal)\in \bR^{t(p+\hat{c})+1}$. It is easy to see that the encoding scheme does not depend on $\cA$ and it is systematic. The storage overhead of the resulting scheme is clearly~$\frac{p+\hat{c}}{p}$. 

    To retrieve a product~$\boldw\boldx^\intercal$, write~$\boldw=(\boldw_1,\ldots,\boldw_t)$ with~$\boldw_i\in\cA^p$ for all~$i$. Let $\theta\triangleq C(\cA)$.
    Using the same arguments as in the proof of Theorem~\ref{theorem:universal}, it can be shown that there exist $\hat{\boldw}^{(1)}, \dots, \hat{\boldw}^{(\theta)} \in \{\pm1\}^k$ and $\{\lambda_i\}_{i=0}^{\theta} \subset \bR$ such that 
    \begin{equation*}
        \boldw\boldx^\intercal= \sum_{i=1}^{\theta}\lambda_i \boldw^{(i)}\boldx^\intercal+\lambda_0\1\boldx^\intercal. 
    \end{equation*}
    For all $i \in [\theta]$, write $\hat{\boldw}^{(i)}=(\boldw^{(i)}_1,\ldots,\boldw^{(i)}_t)$ with each~$\boldw^{(i)}_j\in\{\pm1\}^p$ and note that $\boldw^{(i)}\boldx^\intercal= \sum_{j=1}^{t} \boldw^{(i)}_j\boldx_j^\intercal$. The user can retrieve $\boldw^{(1)}_j\boldx_j^\intercal,\dots, \boldw^{(\theta)}_j\boldx_j^\intercal $ by accessing at most~$R_{\theta}+\theta$ entries from~$\boldy_j$, as follows.
 Since the $\theta$-th generalized covering radius of $\cC$ is $R_{\theta}$, there exist codewords $\boldc_1,\dots,\boldc_{\theta} \in \cC$ such that $ |T| \le R_{\theta}$, where $T=\cup_{i=1}^{\theta} \operatorname{supp}(\boldw^{(i)}_j-\boldc_i)$. Then, by accessing   $\boldc_1\boldx_j^\intercal,\dots,\boldc_{\theta}\boldx_j^\intercal$ and $\{x_{j,u} \vert u \in T\}$,  it is possible to compute ~$\boldw^{(1)}_j\boldx_j^\intercal,\dots, \boldw^{(\theta)}_j\boldx_j^\intercal$. The access requirement for computation of~$\boldw\boldx^\intercal$ is thus at most  $t(R_{\theta}+\theta)+1$, resulting in access ratio of $\frac{R_{\theta}+\theta}{p}$.      
\end{proof}
If a code with $R_{C(\cA)} = C(\cA)R_1$ is used in the above theorem, the resultant $\cA$-feasible pair is the same as that obtained by combining Theorem~\ref{theorem:OLAPimprov} with Theorem~\ref{theorem:universal}. Hamming code $(R_\theta=\theta)$ and entire space $(R_\theta=0)$ are two such codes. The generalized covering radii is a relatively new concept, and it has not been explored in much detail in the literature. The research has been mostly focused on either bounds on generalized covering radius of Reed-Muller codes \cite{elimelech2022generalized, LangtonRaviv}, or asymptotic rate bounds for binary codes given a fixed normalized generalized covering radius \cite{elimelech2021generalized,ElimelechS23}. Asymptotic results are not applicable to our setting, as Theorem~\ref{theorem:generalzied_covering} needs codes with finite block lengths. Additionally, employing Reed-Muller codes in Theorem~\ref{theorem:generalzied_covering} results in feasible pairs that are dominated by those given by Theorem~\ref{theorem:universal}.


We now present an example code construction, found by computer search, which dominates some of the systematic solutions in earlier sections for sets of complexity $2$. Consider the linear code of length $p=9$ with the following generator matrix: 
	\begin{align*}
			\begin{pmatrix}
				1&1&1&1&1&1&1&1&1\\
				0&0&1&0&0&1&1&0&1\\
				0&0&0&1&0&1&0&1&1\\
                0&0&0&0&1&0&1&1&1\\ 
                \end{pmatrix}. 
	\end{align*}
It can be verified that $\hat{c}=8$ and that the second generalized covering radius is~$R_2=3$. Hence, by Theorem~\ref{theorem:generalzied_covering}, the pair $(1.88,0.55)$ is $\cA$-feasible, for all sets $\cA$ of complexity $2$. This pair clearly dominates $(1.89,0.66)$ and $(2.14,0.58)$, obtained by combining the best systematic solutions in Section~\ref{section:two_value} with Theorem~\ref{theorem:universal}. Constructing such good short block-length codes with a small generalized covering radius remains an open and challenging problem and is beyond the scope of this work. 

	\section{Extension to Multivariate Monomial Evaluation} \label{sec:monomials}
 This section shows how the techniques developed in this paper can also be applied to multivariate monomial evaluation. Suppose there exists an $\cA$-protocol for quantized linear computation with parameters $(k, n,\ell)$ and linear decoding. Then, there exist $\{\lambda_{i,j}\}_{i \in [n],j \in [k]} \subset \bR$ such that 
\begin{equation*}y_i=\sum_{j=1}^{k}\lambda_{i,j}x_j,~\forall i \in [n],
\end{equation*} 
where~$\boldy=(y_1,\ldots,y_n)$ is the encoded form of the data~$\boldx=(x_1,\ldots,x_k)$. For any $\boldw \in \cA^k$, there are $\ell$ indices $z_1,\dots,z_{\ell} \in [n]$ such that
 \begin{equation*}\boldw\boldx^\intercal=\sum_{j=1}^{\ell}\mu_{j}y_{z_{j}},
\end{equation*} 
for some $\{\mu_j\}_{j \in [\ell]} \subset \bR$. 
This equation holds irrespective of the value of $\boldx$ and hence,
\begin{equation*}
\sum_{j=1}^{\ell}\mu_{j}\lambda_{z_{j},i}= w_i. 
\end{equation*} 

Now, for a finite set $\cA \subset \bR$, we define the multivariate monomial family with powers restricted to $\cA$ as ~$$\hat{\cF}_{\cA}=\big\{f(\boldx)=\prod_{i=1}^k x_i^{w_i}\vert \boldw \in \cA^k\big\}.$$ 
To avoid corner cases, we only consider evaluations of these monomials on $\boldx$ such that all $x_i$ are non-zero. Here, we are interested in the tradeoff between access-ratio and redundancy-ratio for this family of functions.
We will now construct a protocol for $\hat{\cF}_{\cA}$ with the same $(k,n,\ell)$ parameters as the protocol described above, thereby showing that both $\cF_{\cA}$ and $\hat{\cF}_{\cA}$ have the same feasible pairs.
Let $\hat{\boldy}=(\hat{y}_1,\dots,\hat{y}_n) \in \bR^n$ be defined as
\begin{equation*}\hat{y}_i=\prod_{j=1}^{k}x_j^{\lambda_{i,j}},~\forall i \in [n].
\end{equation*}
Store $\hat{\boldy}$ in a distributed manner over~$n$ servers, with one $\bR$-element~$y_i$ in each server. Then, we have  
\begin{eqnarray*}
        \prod_{i=1}^k x_i^{w_i} &=& \prod_{i=1}^k x_i^{\sum_{j=1}^{\ell}\mu_{j}\lambda_{z_{j},i}} 
    = \prod_{i=1}^k \prod_{j=1}^{\ell} x_i^{\mu_{j}\lambda_{z_{j},i}} \\ &=&  \prod_{j=1}^{\ell} \prod_{i=1}^{k} (x_i^{\lambda_{z_{j},i}})^{\mu_{j}} = \prod_{j=1}^{\ell} \Big(\prod_{i=1}^{k} x_i^{\lambda_{z_{j},i}}\Big)^{\mu_{j}} \\ &=& \prod_{j=1}^{\ell} \hat{y}_{i_j}^{\mu_j}. 
    \end{eqnarray*}
Thus, we have proved that it suffices to access at most $\ell$ entries in $\hat{\boldy}$ to compute any monomial in $\hat{\cF}_{\cA}$. 

\section{Discussion}
 Motivated by applications in machine learning inference, we studied the access-redundancy tradeoff for quantized linear computations.  Using covering codes in a novel way, we developed protocols that outperform the state-of-art for two-valued computations. We then introduced a new additive-combinatorics notion, which we refer to as complexity, and presented a universal protocol for quantized linear computations, for which the redundancy  ratio is independent of the coefficient set and the access ratio depends on the complexity of the coefficient set. We then proceeded to explore this new additive complexity definition in some detail. One key takeaway from this study is that the commonly employed uniform quantization results in sets having the lowest  complexity and thereby leads to a low access ratio. Investigating the existence of efficient algorithms to determine the complexity of a finite set is a challenging open problem.

 The optimality of our techniques remains an unanswered question. The lower bound on the access requirement of $\{\pm1\}$-protocols derived in Section~\ref{section:bound} can be generalized to $\cA$-protocols, giving rise to the following bound.  
	\begin{theorem}\label{theorem:boundgeneral}
		An~$\cA$-protocol with a given~$n$, $k$,~$\ell$, and linear decoding must satisfy that~$\binom{n}{\ell}|A|^\ell\ge |A|^k$.
	\end{theorem} 
	As in Section~\ref{section:bound}, denote~$\nu = \frac{n}{k}$ and~$\lambda = \frac{\ell}{k}$. Let $|\cA|=2^m$. In the large $k$ regime, the bound in Theorem~\ref{theorem:boundgeneral} can be approximated as
	\begin{align}\label{equation:boundgeneral}
		H(\lambda/\nu)\ge \frac{m(1-\lambda)}{\nu},
	\end{align}
	where $H$ is the binary entropy function. We skip the technical details here since it is similar to that in Section~\ref{section:bound}. 
 
 We also note that Proposition~\ref{prop:combination} can be generalized to any finite set $\cA$, resulting in the $\cA$-feasibility of pairs on the line joining any two $\cA$-feasible pairs given by $\cA$-protocols with finite parameters. 
 
  	\begin{figure}[ht!]
  \begin{center}
			\includegraphics[scale=0.5]{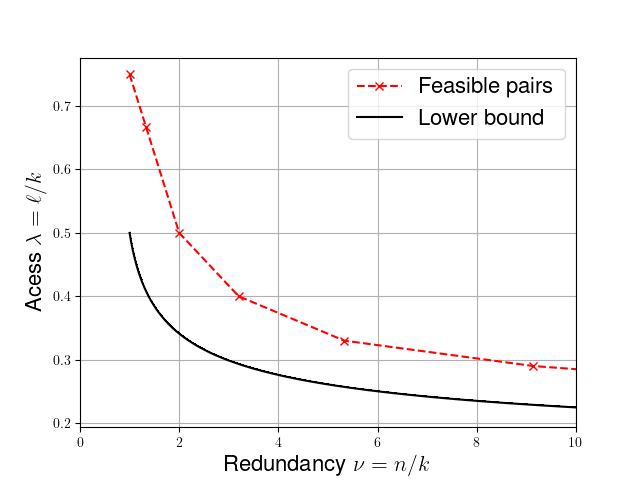}
			\caption{Access-redundancy tradeoff for coefficient set $\cA=\{1,2,3,4\}$.}
			\label{fig:comparison}
		\end{center}
	\end{figure} 
In Fig.~\ref{fig:comparison}, we provide a numerical evaluation of the bound in \eqref{equation:boundgeneral} for any set~$\cA$ of size $4$ and complexity $2$ (e.g.~$\cA=\{1,2,3,4\}$), and compare it against the best feasible pairs obtained using the techniques developed in this paper. It is clear from Fig.~\ref{fig:comparison} that there is a gap between our schemes and bounds. The gap would be even larger if $\cA$ is a set of size $4$ and complexity $3$, as the bound does not depend on the complexity of $\cA$.  

Exploring the possibility of a lower bound which is dependent on the coefficient complexity is an interesting future research direction. Even for the $\{\pm1\}$ case, there is a gap between best constructions and bounds. Developing improved $\{\pm 1\}$-protocols is a prospective research direction.
Another area for future research is further investigating joint computation, with a focus on generalized covering radius. 

 In this paper, we focused on binary covering codes. Given a finite set $\cA$, one could use covering codes over an alphabet of size $|\cA|$ to construct $\cA$-protocols. 
 However, it is not clear if similar gains can be achieved, e.g., via the closure properties mentioned in Observation~\ref{observation:complement}, for an arbitrary covering code.
 The resultant protocol will not be universal, as the encoding scheme will depend on the size of $\cA$. Moreover, the redundancy ratios given by protocols based on large-alphabet covering codes are unlikely to be interesting. Nevertheless, further research is needed to completely understand the applicability of covering codes over larger alphabets, and it is left as future work. 

From a practical standpoint, the number of servers should be substantially smaller than $k$, thereby requiring servers to store multiple real numbers for each data point. Designing such distributed computation systems with an eye toward minimizing both disk access and download is another intriguing future research direction.
		\printbibliography
		\end{document}